\documentclass[10pt,journal,compsoc]{IEEEtran}

\ifCLASSOPTIONcompsoc
  \usepackage[nocompress]{cite}
\else
  \usepackage{cite}
\fi

\ifCLASSINFOpdf
  \usepackage[pdftex]{graphicx}
  \graphicspath{{./pictures/}{./illustrator/}}
  \DeclareGraphicsExtensions{.pdf,.jpeg,.png,.jpg}
\else
\fi
\usepackage{amsmath}
\usepackage{amsopn}
\usepackage{amssymb}
\usepackage{array}
\usepackage{booktabs}                  
\usepackage{color}
\usepackage{esvect}
\usepackage{graphicx}
\usepackage{hyperref}
\usepackage{lipsum}                    
\usepackage{multirow}
\usepackage{mwe}                       
\usepackage{subcaption}
\usepackage{tabu}                      
\usepackage{pdfpages}
\usepackage{wrapfig}
\usepackage{svg}
\usepackage{makecell}

\definecolor{Rosolic}{cmyk}{0.00,1.00,0.50,0}
\definecolor{bleudefrance}{rgb}{0.19, 0.55, 0.91}
\definecolor{cc1}{RGB}{35, 168, 224}
\definecolor{cc2}{RGB}{212, 88, 160}
\definecolor{cc3}{RGB}{245, 184, 47}

\newcommand{\revised}[1]{{#1}}

\usepackage{ulem} 

\newlength{\subht}
\newsavebox{\subbox}

\newcommand{\twosubfig}[5]{
  \sbox\subbox{%
    \resizebox{\dimexpr#1\columnwidth-1em}{!}{%
      \includegraphics[height=3cm]{#3}%
      \includegraphics[height=3cm]{#5}%
    }%
  }
  \setlength{\subht}{\ht\subbox}

  \centering
  \hfill%
  \subcaptionbox{#2}{%
    \includegraphics[height=\subht]{#3}%
  }
  \hfill%
  \subcaptionbox{#4}{%
    \includegraphics[height=\subht]{#5}%
  }
  \hfill%
}

\newcommand{\threesubfig}[7]{
  \sbox\subbox{%
    \resizebox{\dimexpr#1\columnwidth-1em}{!}{%
      \includegraphics[height=3cm]{#3}%
      \includegraphics[height=3cm]{#5}%
      \includegraphics[height=3cm]{#7}%
    }%
  }
  \setlength{\subht}{\ht\subbox}

  \centering
  \hfill%
  \subcaptionbox{#2}{%
    \includegraphics[height=\subht]{#3}%
  }
  \hfill%
  \subcaptionbox{#4}{%
    \includegraphics[height=\subht]{#5}%
  }
  \hfill%
  \subcaptionbox{#6}{%
    \includegraphics[height=\subht]{#7}%
  }
  \hfill%
}

\newcommand{\foursubfig}[9]{
  \sbox\subbox{%
    \resizebox{\dimexpr#1\columnwidth-1em}{!}{%
      \includegraphics[height=3cm]{#3}%
      \includegraphics[height=3cm]{#5}%
      \includegraphics[height=3cm]{#7}%
      \includegraphics[height=3cm]{#9}%
    }%
  }
  \setlength{\subht}{\ht\subbox}

  \centering
  \hfill%
  \subcaptionbox{#2}{%
    \includegraphics[height=\subht]{#3}%
  }
  \hspace{.1mm}%
  \subcaptionbox{#4}{%
    \includegraphics[height=\subht]{#5}%
  }
  \hspace{.1mm}%
  \subcaptionbox{#6}{%
    \includegraphics[height=\subht]{#7}%
  }
  \hspace{.1mm}%
  \subcaptionbox{#8}{%
    \includegraphics[height=\subht]{#9}%
  }
  \hfill%
}

\newcommand{\foursubfignosubcap}[4]{
  \sbox\subbox{%
    \resizebox{\dimexpr\columnwidth-1em}{!}{%
      \includegraphics[height=3cm]{#1}%
      \includegraphics[height=3cm]{#2}%
      \includegraphics[height=3cm]{#3}%
      \includegraphics[height=3cm]{#4}%
    }%
  }
  \setlength{\subht}{\ht\subbox}

  \centering
  \hfill%
  \includegraphics[height=\subht]{#1}%
  \hfill%
  \includegraphics[height=\subht]{#2}%
  \hfill%
  \includegraphics[height=\subht]{#3}%
  \hfill%
  \includegraphics[height=\subht]{#4}%
  \hfill%
}

\newcommand{\foursubfignoindex}[9]{
  \sbox\subbox{%
    \resizebox{\dimexpr#1\columnwidth-1em}{!}{%
      \includegraphics[height=3cm]{#3}%
      \includegraphics[height=3cm]{#5}%
      \includegraphics[height=3cm]{#7}%
      \includegraphics[height=3cm]{#9}%
    }%
  }
  \setlength{\subht}{\ht\subbox}

  \centering
  \hfill%
  \subcaptionbox*{#2}{%
    \includegraphics[height=\subht]{#3}%
  }
  \hspace{.1mm}%
  \subcaptionbox*{#4}{%
    \includegraphics[height=\subht]{#5}%
  }
  \hspace{.1mm}%
  \subcaptionbox*{#6}{%
    \includegraphics[height=\subht]{#7}%
  }
  \hspace{.1mm}%
  \subcaptionbox*{#8}{%
    \includegraphics[height=\subht]{#9}%
  }
  \hfill%
}

\begin{document}
%
\title{Shape Adaptation for 3D Hairstyle Retargeting}
%
%
%
%

\author{Lu~Yu,
        Zhong~Ren,
        Youyi~Zheng,
        Xiang~Chen*,
        and~Kun~Zhou,~\IEEEmembership{Fellow,~IEEE}
\IEEEcompsocitemizethanks{
\IEEEcompsocthanksitem L. Yu, Z. Ren, Y. Zheng, X. Chen and K. Zhou are with the State Key Lab of CAD\&CG, Zijingang Campus, Zhejiang University, Hangzhou, China 310058.}
\thanks{* Corresponding author. E-mail: xchen.cs@gmail.com}}

\IEEEtitleabstractindextext{%
\begin{abstract}
It is demanding to author an existing hairstyle for novel characters in games and VR applications. However, it is a non-trivial task for artists due to the complicated hair geometries and spatial interactions to preserve. In this paper, we present an automatic shape adaptation method to retarget 3D hairstyles. We formulate the adaptation process as a constrained optimization problem, where all the shape properties and spatial relationships are converted into individual objectives and constraints. To make such an optimization on high-resolution hairstyles tractable, we adopt a multi-scale strategy to compute the target positions of the hair strands in a coarse-to-fine manner. The global solving for the inter-strands coupling is restricted to the coarse level, and the solving for fine details is made local and parallel. In addition, we present a novel hairline edit tool to allow for user customization during retargeting. We achieve it by solving physics-based deformations of an embedded membrane to redistribute the hair roots with minimal distortion. We demonstrate the efficacy of our method through quantitative and qualitative experiments on various hairstyles and characters.
\end{abstract}

}

\maketitle

\IEEEdisplaynontitleabstractindextext

%
\IEEEpeerreviewmaketitle

\ifCLASSOPTIONcompsoc
\IEEEraisesectionheading{\section{Introduction}\label{sec:introduction}}
\else
\section{Introduction}
\label{sec:introduction}
\fi

\noindent\textsc{3D hairstyle} is an essential asset for digital characters. In games and VR applications, it is often demanding to author an existing hairstyle for a novel character since the creation of high-quality 3D hairstyles requires laborious effort and expertise \cite{paris2008hair,luo2013structure,shen2023ct2hair}. However, manual retargeting is a non-trivial task as a 3D hairstyle exhibits complicated geometries and spatial interactions, and the artist must ensure a faithful preservation of all these details during authoring (see Figure \ref{fig:overview} b-c).

\begin{figure}[!t]
\centering 
\includegraphics[width=1\columnwidth]{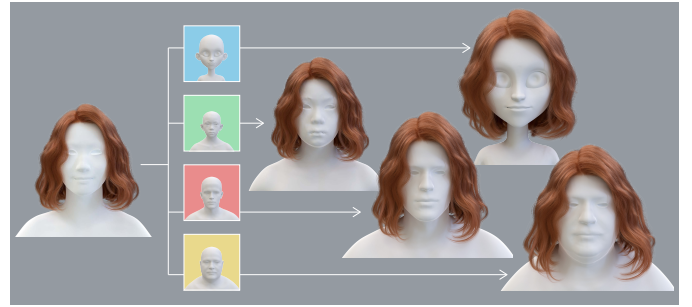}
\caption{Retargeting a 3D hairstyle to different characters.}
\label{fig:teaser}
\end{figure}

In this paper, we present an automatic shape adaptation method for 3D hairstyle retargeting. Given the hairstyle of a source character, our method can adapt it to diverse targets with distinct body shapes (see Figure \ref{fig:teaser}). Although there are 2D hairstyle transfer methods like \cite{tan2020michigan,zhu2021barbershop,kim2022style} investigated with the recent progress of StyleGAN \cite{karras2019style,karras2020analyzing}, the pixel-level manipulations cannot be applied to 3D hairstyles. As a 3D hairstyle possesses various per-strand shapes, complicated inter-strand relationships, and complex interactions with the character, a key challenge is how to preserve such geometric information in accordance with the shape variance between the source and target characters. Therefore, we formulate the adaptation process as a constrained optimization, where all the shape properties and spatial relationships are converted into individual objectives and constraints. 

Shape adaptation can be computationally intractable for 3D hairstyles. In order to express sufficient fine details, a 3D hairstyle often possesses hundreds of thousands of strands and millions of particles. With such a high-resolution model, the global coupling between all hair strands leads to a huge system to solve. To tackle this problem, we present a multi-scale solving strategy to organize the hair strands into two levels of detail and restrict global solving to the coarse level. With the coarse adaptation ready, solving the fine details is made local and parallel.

We further allow the users to customize the position and shape of the hairline to obtain more variants. A critical step to support hairline edits is redistributing the hair roots by stretching a scalp membrane embedded in the head surface. To minimize the distortion of such a redistribution, we resort to a physics-based formulation, i.e., optimizing an elasticity energy parameterized intrinsically in the surface space.

We conduct experiments on various hairstyles and characters. The qualitative and quantitative results demonstrate the fidelity and diversity of our retargeting method. Further-more, various hairline edits from users are supported, and the multi-scale solving achieves a two-orders-of-magnitude speedup, compared with the full solving of all strands.

Our main contributions are:
\begin{itemize}
\item The first shape adaptation framework based on constrained optimization for 3D hairstyle retargeting;
\item A multi-scale solving method to enable the optimization of high-resolution hairstyles;
\item A physics-based embedded membrane deformation method to support hairline edits.
\end{itemize}

\section{Related Work}
\subsection{3D Hair Modeling}
With the increasing demand for realistic hairstyles in digital humans, 3D hair modeling techniques have been continuously improved \cite{ward2007survey,bao2018survey}. High-quality reconstruction of 3D hairstyles often requires a complex acquisition setup, e.g., the LED light sources, robotic arms, SLR camera arrays \cite{paris2008hair,luo2013structure}, or even computed-tomography scanners\cite{shen2023ct2hair}. The obtained hairstyles are of satisfying fidelity, but the expensive equipment or long capturing time restricts the acquisition scalability. Recently, data-driven and deep learning methods are adopted to generate 3D hairstyles from single-view \cite{hu2015single,chai2016autohair,saito20183d,wu2022neuralhdhair} or multi-view images \cite{zhang2017data,kuang2022deepmvshair}, and the depth information is also used \cite{hu2014capturing,zhang2018modeling}. These image-based methods are straightforward to scale up. However, the produced 3D hairstyles cannot yet be compared with the acquisition results in terms of precision and quality, mainly due to the limited 3D hair models for training and the inherent domain gap between the real images and the synthetic ones.

Generative models for 3D hairstyles have been recently developed. The variational autoencoder is often employed to learn the representation model for individual strands and hairstyles. Hairstyle variants can be efficiently synthesized through a latent-space sampling or interpolation \cite{zhou2023groomgen}. Text conditioning to a diffusion network is also enabled to help guide the hairstyle generation \cite{sklyarova2024text}. These works conduct the training on a canonical head shape that would be implicitly embedded within the network, and the synthesis is bound to this shape inherently. Our method complements them since we can effectively retarget the ones they produce to novel characters, vastly increasing the variants.

\subsection{2D Hairstyle Transfer}
Given the data scarcity of high-quality 3D hairstyles, reusing existing models has practical significance. Specifically, methods are investigated for hairstyle transfer in the image space \cite{tan2020michigan,saha2021loho,zhu2021barbershop,kim2022style}. MichiGAN \cite{tan2020michigan} designs individual condition modules to compose the shape, structure, appearance, and background attributes of different source images into a new one. Differently, LOHO \cite{saha2021loho} optimizes the latent space of StyleGAN2 \cite{karras2020analyzing} with a gradient orthogonalization strategy to disentangle the hair attributes. Based on the latent space of StyleGAN2, Barbershop \cite{zhu2021barbershop} enhances its spatial encoding with a novel structure tensor and adopts a semantic alignment step to improve the blending coherence. StyleYourHair \cite{kim2022style} employs facial-keypoints detection and a dedicated local-style-matching loss to deal with the pose difference between the reference images. Those methods do not apply directly to 3D hairstyle transfer. First, the latent-space hair manipulation depends on an underlying representation model. This necessitates a large training set that is publicly available to facial images but not 3D hair models. Second, 2D hairstyle transfer focuses on the visual plausibility of the pixel-space compositions under a given viewpoint, while a 3D method explicitly adapts the hair geometries. During the adaptation, we must ensure the full fidelity of hair shapes and preserve their spatial interactions with the characters. Moreover, our method does not rely on any 3D hair dataset. In fact, it could be leveraged to vary existing models for data augmentation.


\subsection{3D Garment Retargeting}
Although there has yet to be direct research on 3D hairstyle transfer, to the best of our knowledge, a few methods have been developed for resizing the garments to fit a target body. Earlier methods adopt skinning-type strategies to warp the garment surface mesh along with the character body \cite{cordier2003made,wang2005design,wang2007volume}. The warping result is tightly bound to the shape variance of the body mesh, leading to undesirable artifacts for loose garments. Then, Meng et al. \cite{meng2012flexible} propose manually adding user guidances to regularize the garment shape, but merely using the annotations on sparse views is insufficient to eliminate all the shape artifacts. Brouet et al. \cite{brouet2012design} presents a gradient-based objective and an operation for deformation projection to explicitly preserve the shape properties during garment transfer. The techniques from these methods could be borrowed to solve transfer problems for other assets. For example, most of them have defined the spatial relationship between the garment and body, though using diverse ways, and we modify the skeleton-based local positioning strategy presented in \cite{brouet2012design} to express the hair-body relationship. The methods are not fully applicable to hairstyle transfer due to the significant differences between the geometric representations of hairstyle and garment. The garment is modeled as a set of connected 2D surface meshes, while the hairstyle is composed of many individual 1D strands, all embedded in the 3D space. In this work, we design dedicated energies to model per-strand shapes and inter-strand interactions for hairstyle retargeting. Furthermore, we present a multi-scale solving strategy to deal with the high-resolution nature of hair models in a coarse-to-fine manner, as well as a hair-root relocation algorithm to support the user edits of hairline.

Recently, data-driven and learning-based methods have been utilized to retarget the pattern-based garment designs \cite{wang2018learning}, cloth dynamics \cite{guan2012drape,santesteban2019learning} and wrinkles \cite{pons2017clothcap,lahner2018deepwrinkles}. These methods often require a training dataset with paired combinations of garment samples over body samples. Our method can be an optional tool to create such a dataset for hair.

\begin{figure*}[!t]
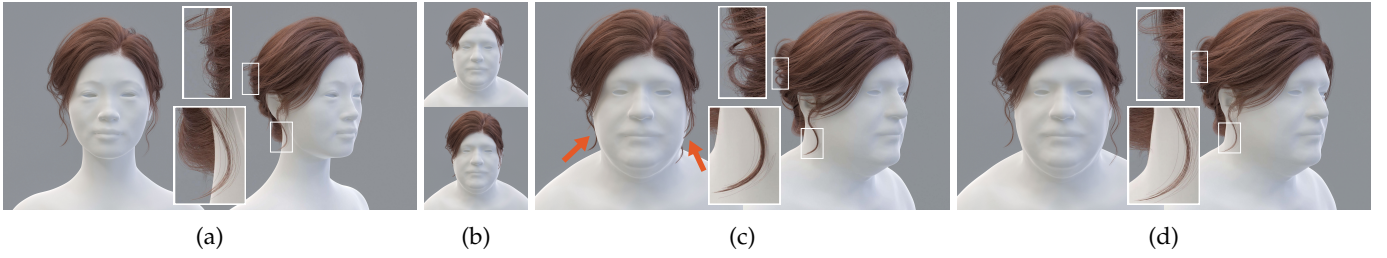

\centering
\begin{minipage}{1.0\textwidth}
\foursubfig{1.0}
	{}{overview/source}
	{}{overview/rigid}
	{}{overview/affine}
	{}{overview/our_transfer}
\end{minipage}
\caption{(a) The source character and hairstyle. (b) A naive retargeting, using either a global translation (up) or per-strand translations based on the barycentric coordinates of hair roots on the scalp (bottom), produces wrong proportions between the hair and the character. (c) With a global affine transformation and per-strand translations, the large-scale proportion improves. However, the retargeted hairstyle still has the wrong hair-body relationship (e.g., see the arrows: the tight attaching and interpenetration artifacts between the side fringes and the cheek) and inter-strand relationship (e.g., see the zoom-ins: the hallucinated and exaggerated bundle concentrations). (d) Our method produces a faithful retargeting result exhibiting, e.g., side fringes with reasonable gaps to the cheek and bundles with accurate, fluffy shapes.}
\label{fig:overview}
\end{figure*}

\section{Overview}
\label{section:Overview}
We aim to retarget strand-based 3D hairstyles.
Specifically, given two characters with similar poses, our method takes an arbitrary hairstyle customized for the source and adapts its shape to the target. The produced hairstyle preserves a consistent look with the original hair while being compatible with the target character.

Due to the inherent shape differences between the source and the target character, naively transferred hairstyles often possess one or many of the following issues.

The \textit{single-strand shape} is visually crucial to 3D hairstyles. Without any constraints for it, a position-based optimization on the strand particles would easily damage the per-strand shape properties, e.g., curliness.

The \textit{inter-strand relationship} represents the spatial interactions between different strands in a hairstyle. Hair strands are topologically independent, but their relative locations in 3D are visually salient. Thus, processing each strand without any mutual constraints would quickly lose this global shape information (see insets of Figure \ref{fig:overview}c).

The \textit{hair-body relationship} describes the relative positions between the hair strands and the character model, especially the head, neck, shoulder, and upper arms. Such spatial relationships could be inherently semantic, e.g., a side fringe beside the cheek. To faithfully capture these abundant mutual semantics, solely considering the hair-roots positioning on the scalp during transfer is far from satisfactory (Figure \ref{fig:overview}b). Things get even worse when the transferred hairs \textit{penetrate} the body, which leads to severe artifacts (Figure \ref{fig:overview}c).

The above shape factors could conflict; for example, strict preservation of the hair-body relationships could harm the single-strand shapes. Thus, the retargeting method should sensibly balance those factors and produce a hairstyle that is visually faithful to the source models and spatially adaptive to the target character, e.g., penetration-free. Meanwhile, the method must be fast to compute and flexible for user edits.

To this end, we carefully design a two-stage method for 3D hairstyle retargeting (Section \ref{sec:hairstyle_transfer_method}). First, we compute an initial hairstyle that preserves the hair-body relationship as much as possible but ignores all the other shape factors. Second, starting from the initial transfer result, we simultaneously optimize all the mentioned shape factors by solving an iterative quadratic programming problem to obtain the final hairstyle.

To accelerate the optimization, we build a hair hierarchy to decouple the inter-strand relationships, enabling approximate but efficient multi-scale solving (Section \ref{sec:mss}). To support user edits of the hairline, we present a physics-based embedded membrane deformation to redistribute the hair roots on the target scalp with minimal distortion (Section \ref{sec:hairline_change}).

%


\section{3D Hairstyle Retargeting}
\label{sec:hairstyle_transfer_method}
We represent a 3D hairstyle as independent strands, where each strand is composed of sequentially connected particles. We stack per-particle 3D positions of the original and the transformed hairstyle to vectors $\mathbf{\bar{p}},\mathbf{p}\in\mathbb{R}^{3\times N}$, where $N$ is the total number of particles. We denote the source and target character mesh by $M_\mathrm{s}$ and $M_\mathrm{t}$, and their 3D positions by $\mathbf{\bar{q}}$ and $\mathbf{q}$, respectively. Moreover, we assume that $M_\mathrm{s}$ and $M_\mathrm{t}$ have the same topology.

For the 3D hairstyle retargeting, we seek to find a good $\mathbf{p}$ given $\mathbf{\bar{p}}$, $\mathbf{\bar{q}}$ and $\mathbf{q}$. Mathematically, we formulate this \textit{shape adaptation} as a constrained nonlinear optimization problem.
\begin{equation}
\label{eqn:constrained_optimization}
\begin{split}
\min_\mathbf{p} \quad & E_\mathsf{strand\text{-}shape} + \alpha E_\mathsf{inter\text{-}strand} + \beta E_\mathsf{hair\text{-}body} \\
\mathrm{s.t.} \quad & C_\mathsf{root},C_\mathsf{penetration}
\end{split}
\end{equation}

\subsection{\revised{Initial Transfer}}
\label{sec:initial_transfer}


The nonlinear optimization necessitates an initial value of $\mathbf{p}$. In order to stabilize and accelerate the convergence, a good initial hairstyle should possess a smooth shape and correct proportions. Thus, we compute such a value by conforming to the source hair-body relationship.

We first generate the source and target skeleton $K_\mathrm{s}$ and $K_\mathrm{t}$, with the same topology, by fitting the SMPL \cite{loper2015smpl} to the character models $M_\mathrm{s}$ and $M_\mathrm{t}$, respectively.

Next, we adapt the cloth positioning strategy \cite{brouet2012design} to localize hairs. For each particle $\mathbf{\bar{p}}_{i}$ of the source hairstyle, we choose a bone from $K_\mathrm{s}$ as the anchor for local positioning:
\begin{equation}
\label{eqn:bone_indicator}
\min_{{b}\in\mathrm{bones}} \mathbf {1}_{\mathbf{q}_{b}\in\Lambda_{b}} \cdot  \left\|\mathbf{\bar{p}}_{i}-\mathbf{\bar{q}}_{b}\right\| \cdot e^{\sigma\left<\mathbf{\bar{r}}_{b},\mathbf{\bar{v}}_{b}\right>^2}.
\end{equation}
As shown in \revised{Figure \ref{fig:ss_rel_and_hb_rel}a}, on each bone we find the point $\mathbf{\bar{o}}_{b}$ that is closest to $\mathbf{\bar{p}}_{i}$, and we find the intersection between the ray $\vv{\mathbf{\bar{o}}_{b}\,\mathbf{\bar{p}}_{i}}$ and the source character $M_\mathrm{s}$ as the point $\mathbf{\bar{q}}_{b}$. In the above exponentiation, $\mathbf{\bar{r}}_{b}=\frac{\mathbf{\bar{q}}_{b}-\mathbf{\bar{o}}_{b}}{\left\|\mathbf{\bar{q}}_{b}-\mathbf{\bar{o}}_{b}\right\|}$ is the direction of the ray, $\mathbf{\bar{v}}_{b}$ is the direction of the bone, and $\sigma$ is a constant that we set to $100$. In other words, we tend to choose a bone orthogonal to the ray with an intersection point near the hair particle. As this simple metric \cite{brouet2012design} may choose a faraway bone with false intersection, e.g., a ray starting from clavicle intersects the scalp, we add an indicator function $\mathbf {1}_{\mathbf{\bar{q}}_{b}\in\Lambda_{b}}$ to ensure that the intersection point is within a valid surface region. We clip the skinning weights of SMPL for bone $\mathrm{b}$ to determine this valid region $\Lambda_{b}$.

For each $\mathbf{\bar{p}}_{i}$, we record the length $\bar{\eta}_{i} = \left\|\mathbf{\bar{p}}_{i}-\mathbf{\bar{q}}_{b}\right\|$ as the local coordinates. We also use the barycentric coordinates of $\mathbf{\bar{q}}_{b}$ and $\mathbf{\bar{o}}_{b}$ to find the reference points $\mathbf{q}_{b}$ and $\mathbf{o}_{b}$ on target models $K_\mathrm{t}$ and $M_\mathrm{t}$. By applying the local coordinates to the reference points, we obtain the target position of particle $\mathbf{\bar{p}}_{i}$:
\begin{equation}
\label{eqn:hair_body_recover}
\mathbf{\tilde{p}}_{i} = \mathbf{q}_{b} + \bar{\eta}_{i} \cdot \frac{\mathbf{q}_{b}-\mathbf{o}_{b}}{\left\|\mathbf{q}_{b}-\mathbf{o}_{b}\right\|}.
\end{equation}

The position $\mathbf{\tilde{p}}_{i}$ is a local retargeting due to its individual computation on each particle. This leads to stairstep strands, which hinders the optimization in Equation \ref{eqn:constrained_optimization}. Therefore, we smooth each strand by solving a discrete Poisson's equation for $\mathbf{\hat{p}}$ with the Dirichlet boundary condition:
\begin{equation}
\label{eqn:poisson_smooth}
\begin{split}
\mathbf{L}_{i}[\mathbf{\hat{p}}] = \mathbf{L}_{i}[\mathbf{\bar{p}}], \quad &\forall i\in \mathcal{I}_\mathrm{u} \\
\mathrm{s.t.} \quad \mathbf{\hat{p}}_{j} = \mathbf{\tilde{p}}_{j}, \quad &\forall j\in \mathcal{I} - \mathcal{I}_\mathrm{u},
\end{split}
\end{equation}
where $\mathbf{L}[\ast]$ is a discrete Laplace operator defined on $\mathbf{\bar{p}}$:
\begin{equation}
\label{eqn:strand_laplacian}
\mathbf{L}_{i}[f] = \frac{f_{i+1}-f_{i}}{\left\|\mathbf{\bar{p}}_{i+1}-\mathbf{\bar{p}}_{i}\right\|} - \frac{f_{i}-f_{i-1}}{\left\|\mathbf{\bar{p}}_{i}-\mathbf{\bar{p}}_{i-1}\right\|},
\end{equation}
and \revised{$\mathcal{I}_\mathrm{u}$ is the index set of the particles with large Laplacian feature discrepancies from the source:}
\begin{equation}
\label{eqn:sharp_corners}
\mathcal{I}_\mathrm{u} = \{\, {i}\, \mid \,\left\|\mathbf{L}_{i}[\mathbf{\tilde{p}}] - \mathbf{L}_{i}[\mathbf{\bar{p}}]\right\|_2 > \epsilon_\mathrm{s}\,\}.
\end{equation}

\subsection{Single-Strand Shape}
In Equation \ref{eqn:constrained_optimization}, the first energy term measures how large the strand shapes change during retargeting. At first glance, it is appealing to define the local shape change as the variation of individual strand segments, where each segment connects two adjacent particles in a strand. However, the retargeting often necessitates a proportion change due to the significant shape difference between the source and target character. To this end, we allow a segment to change length but penalize its change in direction:
\begin{equation}
\label{eqn:single_strand_shape}
E_\mathsf{strand\text{-}shape} = \sum_{{a,b}\in\mathrm{segments}} \left\|\frac{\mathbf{p}_{a}-\mathbf{p}_{b}}{\left\|\mathbf{p}_{a}-\mathbf{p}_{b}\right\|_2}- \frac{\mathbf{\bar{p}}_{a}-\mathbf{\bar{p}}_{b}}{\left\|\mathbf{\bar{p}}_{a}-\mathbf{\bar{p}}_{b}\right\|_2}\right\|^2_2.
\end{equation}

\subsection{Inter-Strand Relationship}
\label{sec:strand_strand_relationship}
\begin{figure}[!t]
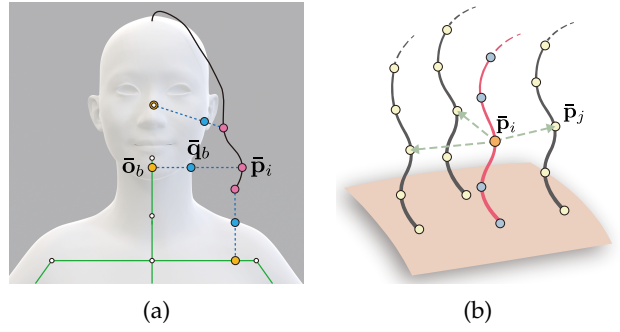

\twosubfig{.9}
	{}{hair-body-relationship/hair-body-relationship}
    {}{hair-hair-relationship/hair-hair-relationship}
\caption{\revised{(a) The geometric references for computing the local positioning coordinates of the hair-body relationship. (b) The nearest particles for computing the Laplacian features of the inter-strand relationship.}}
\label{fig:ss_rel_and_hb_rel}
\end{figure}

The second energy term in Equation \ref{eqn:constrained_optimization} measures the change of spatial relationships between different strands. For each hair particle in the source hairstyle, we find the $k$ particles nearest to it in different strands and then build a distance-weighted Laplacian feature to encode the relative positions between them (see \revised{Figure \ref{fig:ss_rel_and_hb_rel}b}):
\begin{equation}
\label{eqn:laplacian_feature}
\mathbf{\bar{L}}_{i} = \sum_{{j}\in\mathrm{knn}(i)} \omega_{j}\left(\mathbf{\bar{p}}_{i}-\mathbf{\bar{p}}_{j}\right).
\end{equation}

Considering that closer points imply stronger bindings, we compute the inverse of the distances and normalize them to obtain the Laplacian weights:
\begin{equation}
\label{eqn:laplacian_weight}
\sigma_{j} = \frac{1}{\|\mathbf{\bar{p}}_{i}-\mathbf{\bar{p}}_{j}\|}, \quad \omega_{j} = \frac{\sigma_{j}}{\sum_{\scriptscriptstyle{m}\in\mathrm{knn}(i)} \sigma_{m}}.
\end{equation}

Using the nearest neighbors and Laplacian weights computed above, we generate the same feature for the adapted hairstyle and penalize its change during retargeting:
\begin{equation}
\label{eqn:strand_strand_relationship}
\mathbf{L}_{i} = \sum_{{j}\in\mathrm{knn}(i)} \omega_{j}\left(\mathbf{p}_{i}-\mathbf{p}_{j}\right), \quad E_\mathsf{inter\text{-}strand} = \sum_{i=0}^N \left\|\mathbf{L}_{i} - \mathbf{\bar{L}}_{i}\right\|_2^2.
\end{equation}

\subsection{Hair-Body Relationship}
In the first stage of our method (see Section \ref{sec:initial_transfer}), we produce an initial transfer result $\mathbf{\hat{p}}$ according to the spatial relationship between the hairstyle and the character model. Besides serving as the initial value for optimization, we also use $\mathbf{\hat{p}}$ to regularize the optimization variable $\mathbf{p}$ in the third energy term of Equation \ref{eqn:constrained_optimization}:
\begin{equation}
\label{eqn:hair_body_relationship}
E_\mathsf{hair\text{-}body} = \sum_{i=0}^N \left\|\mathbf{p}_{i} - \mathbf{\hat{p}}_{i}\right\|_2^2
\end{equation}

\subsection{Constraints}
After the initial transfer, we obtain the position on the target scalp where each strand should grow from, and we fix each root particle to stay at this position:
\begin{equation}
\label{eqn:root_constraint}
C_\mathsf{root}: \quad \mathbf{p}_{i} = \mathbf{\hat{p}}_{i}, \quad \forall i\in \mathcal{I}_\mathrm{r},
\end{equation}
where $\mathcal{I}_\mathrm{r}$ is the index set of root particles.

To ensure the transfer is penetration-free, we project each non-root particle $\mathbf{p}_{i}$ onto the target character model $M_\mathrm{t}$ to get the projection point $\mathbf{q}_{i}$ and the surface normal $\mathbf{n}_{i}$ at this point. Then we constrain the particle $\mathbf{p}_{i}$ to stay in the upper half-space defined by $\mathbf{q}_{i}$ and $\mathbf{n}_{i}$:
\begin{equation}
\label{eqn:penetration_constraint}
C_\mathsf{penetration}: \quad \left< \mathbf{p}_{i}-\mathbf{q}_{i},\, \mathbf{n}_{i} \right> \geq \epsilon_\mathrm{c}, \quad \forall i\in \mathcal{I}_\mathrm{p},
\end{equation}
where $\epsilon_\mathrm{c}$ is a safe-guard for clearance, and $\mathcal{I}_\mathrm{p} = \mathcal{I} - \mathcal{I}_\mathrm{r}$ is the index set of non-root particles.

\subsection{Optimization}
\label{sec:optimization}
The complexity of Equation \ref{eqn:constrained_optimization} stems from two places. One is the nonlinear denominator $\left\|\mathbf{p}_{a}-\mathbf{p}_{b}\right\|_2$ in $E_\mathsf{strand\text{-}shape}$. The other is the dependency of $\mathbf{q}_{i}$ and $\mathbf{n}_{i}$ on $\mathbf{p}_{i}$ in $C_\mathsf{penetration}$. To solve the optimization in a robust and tractable manner, we adopt an iterative update strategy.

In each iteration, we first use the current values of the optimization variable $\mathbf{p}$ to compute each value of $\left\|\mathbf{p}_{a}-\mathbf{p}_{b}\right\|_2$, $\mathbf{q}_{i}$, and $\mathbf{n}_{i}$. Then we fix these values to transform the original formulation of Equation \ref{eqn:constrained_optimization} into a classic quadratic programming problem, since now all the objectives are quadratic and all the constraints are linear, with respect to $\mathbf{p}$. Specifically, we use the ADMM algorithm \cite{boyd2011distributed} to solve this QP problem and update $\mathbf{p}$. We end the iterations when $\mathbf{p}$ converges.

\section{Multi-scale Solving}
\label{sec:mss}
For a typical 3D hairstyle with hundreds of thousands of strands, the Laplacian energy in Equation \ref{eqn:strand_strand_relationship} has to deal with the global coupling between millions of unknown variables. This requires solving a large sparse linear system, which is the bottleneck of the optimization in Section \ref{sec:optimization}.

\begin{figure}[!t]
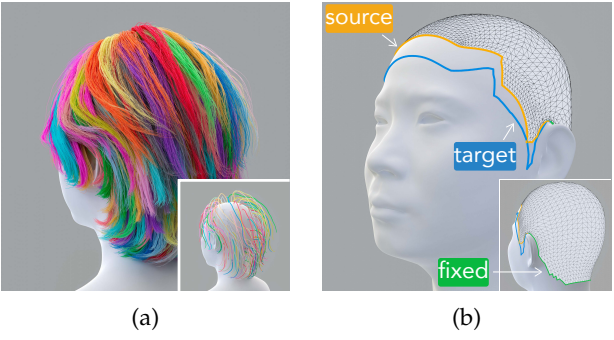

\twosubfig{.9}
	{}{hair_clustering/hair_clustering.pdf}	
        {}{hairline_change/hairline_change.pdf}
\caption{(a) Hair clustering for multi-scale solving. The inset shows the guide hairs. (b) The front part of the source and target hairlines. The inset shows the fixed part.}
\label{fig:hair_clusters_and_hairline_change}
\end{figure}

To accelerate the transfer optimization, we build a hierarchy for the hair strands and separate the computation into two scales. \revised{A similar strategy is also used in recent studies for real-time hair simulations \cite{chai2014reduced,lyu2020real,hsu2024real}. Specifically, we apply the clustering method in \cite{wang2009example} to the original hairstyle to choose a small set of representative hair strands, i.e., the \textit{guide} hairs (Figure \ref{fig:hair_clusters_and_hairline_change}a). The rest of the hair strands are called \textit{normal} hairs.}

In the coarse scale, we adapt the guide hairs altogether to establish the global shape structure of the retargeting. We use the same strategy introduced in Section \ref{sec:optimization} to optimize Equation \ref{eqn:constrained_optimization}, but only for the guide hairs. Since their number is quite low, the sparse linear system from the inter-strands coupling becomes much smaller. After this global solving, all the guide hairs have already been adapted to the target character while preserving their shape features and spatial interactions.

In the fine scale, we fix all the adapted guide hairs and use them to constrain the adaptation of normal hairs. To this end, we adjust the Laplacian defined in Section \ref{sec:strand_strand_relationship} to:
\begin{equation}
\label{eqn:decoupled_laplacian_feature}
\mathbf{\bar{L}}_{i} = \sum_{{j}\in\mathrm{knn}(i,\,\mathcal{I}_\mathrm{g})} \omega_{j}\left(\mathbf{\bar{p}}_{i}-\mathbf{\bar{p}}_{j}\right), \quad \forall i\in \mathcal{I}_\mathrm{n},
\end{equation}
where $\mathcal{I}_\mathrm{g}$ is the index set of the guide hair particles and $\mathcal{I}_\mathrm{n} = \mathcal{I} - \mathcal{I}_\mathrm{g}$ is of the normal hair particles. In other words, we find the nearest particles of each normal hair particle \textit{only from} the guide hairs, and use these neighbors to compute the Laplacian. This adjustment breaks the global coupling between the normal strands, and each strand only depends on its nearby guide strands. As a result, we separate a large system into independent parts for individual strands, which can be readily solved in parallel. This decoupling is based on the local coherency of hairstyles and is a local approximation in essence, which is the key to a significant acceleration. The users could balance the transfer speed and fidelity by tuning the number of guide hairs. \revised{We optimize Equation \ref{eqn:constrained_optimization} with this decoupled Laplacian to achieve the normal hairs adaptation.}

\section{Hairline Edit}
\label{sec:hairline_change}
During hairstyle transfer, users often require design changes for customization. We introduce a method to support user-edits of the hairline, as it is one of the salient features of a hairstyle. To enable this, we first relocate all the hair roots on the target head to obey the user-input hairline by solving a physics-based deformation model, and then we adjust the objectives in Equation \ref{eqn:constrained_optimization} to accommodate the relocated hair roots and fine-tune the optimization.



\subsection{Hair Roots Relocation}
\label{sec:hair_roots_relocation}
After the initial transfer, we obtain the positions of the hair roots on the target head. Then, we extract its scalp region by collecting all the head mesh triangles with at least one root particle inside. The boundary of the scalp, i.e., the \textit{hairline}, is separated into front/back parts at the ears. We fix the back part and let users draw a curve to specify the desired look of the front part (Figure \ref{fig:hair_clusters_and_hairline_change}b). We also extract the turning points of the hairline and let the user specify their corresponding positions on the drawing curve. The dense correspondence for each segment between adjacent turning points is found by mapping the normalized arc-length parameters.

With the current shape of the scalp $\mathcal{S}$ as the undeformed configuration and the hairline correspondence $h(\ast)$ as the boundary condition, we compute the target configuration of $\mathcal{S}$ by solving a deformation problem. As a key insight, the stretching and compression of the scalp should be restricted to the surface space of the head. Therefore, we treat $\mathcal{S}$ as a membrane model embedded in the head surface and minimize a hyperelastic strain energy under Dirichlet boundary condition:
\begin{equation}
\label{eqn:membrane_on_head}
\begin{split}
\min_\mathbf{u} \quad&  \sum_{{t}\in\mathcal{T}_\mathcal{S}} A_{t}\cdot\psi\left(\mathbf{F}_{t}(\mathbf{x}(\mathbf{u}),\mathbf{X})\right) \\
\mathrm{s.t.} \quad & \mathbf{x}_{i} = h(\mathbf{X}_{i}), \quad \forall {i} \in \mathcal{I}_{\partial\mathcal{S}}.
\end{split}
\end{equation}
Here $\mathcal{T}_\mathcal{S}$ is the index set of the triangles in $\mathcal{S}$, and $A_{t}$ is the area of triangle ${t}$, whose deformation gradient $\mathbf{F}_{t}$ depends on the undeformed positions $\mathbf{X}$ and deformed positions $\mathbf{x}$ of the vertices in $\mathcal{S}$. To enable the head surface embedding, we rely on an arbitrary parameterization space, e.g., \cite{eck1995multiresolution}, of the head surface $\mathcal{H}$ to conduct the optimization. The variable to optimize is $\mathbf{u}$, i.e., $\mathbf{x}$'s counterpart in parameterization space. We choose the neo-Hookean membrane model as the strain energy density function $\psi$, which takes $\mathbf{F}$ as the input. The index set $\mathcal{I}_{\partial\mathcal{S}}$ contains all the hairline vertices.

\begin{figure}[!ht]
\centering 
\includegraphics[width=1\columnwidth]{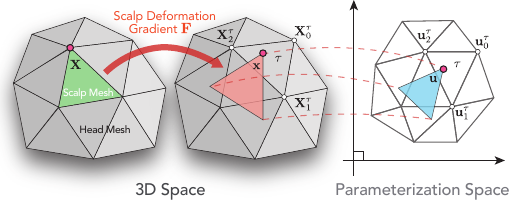}
\caption{The scalp region is a subset of the head mesh. We copy it as the undeformed configuration and constrain the scalp mesh to deform \textit{only on} the head mesh. The energy in Equation \ref{eqn:membrane_on_head} measures how large the scalp deformation is in the 3D space. Meanwhile, the deformed 3D position $\mathbf{x}$ of a scalp vertex is determined by its 2D coordinates $\mathbf{u}$, which we optimize in Equation \ref{eqn:membrane_on_head}, in the head parameterization space. In this way, we model the scalp as an embedded membrane.}
\label{fig:hairline_dg_ref}
\end{figure}

The deformation gradient $\mathbf{F}_{t}$ is a constant $3 \times 2$ matrix for each scalp triangle ${t}$. Now we omit the subscript ${t}$ and represent a deformation gradient as $\mathbf{F} = \mathbf{d}\cdot\mathbf{D}^{-1}$, where $\mathbf{D}$ is a $2\times2$ matrix stacking two adjacent edge vectors of the undeformed triangle in a reparameterized 2D space:
\begin{equation}
\label{eqn:dg_undeformed_part}
\begin{split}
\mathbf{D} &= \left[\mathbf{\hat{X}}_1 - \mathbf{\hat{X}}_0 \mid \mathbf{\hat{X}}_2 - \mathbf{\hat{X}}_0\right] \\
&= \begin{bmatrix}
        \|\mathbf{X}_1-\mathbf{X}_0\| & \frac{(\mathbf{X}_2-\mathbf{X}_0)\cdot(\mathbf{X}_1-\mathbf{X}_0)}{\|\mathbf{X}_1-\mathbf{X}_0\|}\\[1.5mm]
        0 & \frac{\|(\mathbf{X}_2-\mathbf{X}_0)\times(\mathbf{X}_1-\mathbf{X}_0)\|}{\|\mathbf{X}_1-\mathbf{X}_0\|}\\
    \end{bmatrix},
\end{split}
\end{equation}
and $\mathbf{d}$ is a $3\times2$ matrix stacking the two edge vectors of the deformed triangle in the 3D space:
\begin{equation}
\label{eqn:dg_deformed_part}
\mathbf{d} = \left[\mathbf{x}_1 - \mathbf{x}_0 \mid \mathbf{x}_2 - \mathbf{x}_0\right].
\end{equation}
Since the scalp is required to deform on the head surface, a single deformed vertex of $\mathcal{S}$ moves locally inside a triangle $\tau$ of $\mathcal{H}$, whose 3D coordinates $\mathbf{x}$ is a barycentric interpolation on vertices of triangle $\tau$ (see Figure \ref{fig:hairline_dg_ref}):
\begin{equation}
\label{eqn:x_barycentric_interp}
\mathbf{x} = \left[\mathbf{X}^\tau_1 - \mathbf{X}^\tau_0 \mid \mathbf{X}^\tau_2 - \mathbf{X}^\tau_0\right]\lambda+ \mathbf{X}^\tau_0,
\end{equation}
where $\lambda$ is the 2D barycentric coordinates computed in the parameterization space of $\mathcal{H}$ for the intrinsic parameters $\mathbf{u}$ corresponding to $\mathbf{x}$:
\begin{equation}
\label{eqn:u_barycentric_solve}
\lambda = \left[\mathbf{u}^\tau_1 - \mathbf{u}^\tau_0 \mid \mathbf{u}^\tau_2 - \mathbf{u}^\tau_0\right]^{-1}\left(\mathbf{u} - \mathbf{u}^\tau_0\right).
\end{equation}
According to Equations \ref{eqn:dg_undeformed_part}-\ref{eqn:u_barycentric_solve}, $\mathbf{F}$ is a linear function of $\mathbf{u}$.

We solve for the 2D parameters $\mathbf{u}$ of each deformed scalp vertex by minimizing the nonlinear objective in Equation \ref{eqn:membrane_on_head} under the constraints from hairline correspondence $h(\ast)$. In practice, we conduct this optimization using the Projected Newton method \cite{teran2005robust} with a backtracking line search. The host triangle indices $\tau$ must be updated after each Newton iteration according to the current values of $\mathbf{u}$, i.e., the vertex positions in the parameterization space. We use a harmonic map \cite{eck1995multiresolution} to parameterize $\mathcal{H}$, though the converged result is insensitive to the choice of parameterization (Figure \ref{fig:hc_diff_param_neo_hookean}). Besides, our embedded membrane resembles the Lagrangian-on-Lagrangian approach \cite{montes2020computational} that employs the limit geometry of the Loop subdivision to eliminate the discontinuities across triangles. For our problem, the iterative solving in a single parameterization space of the surface is sufficient.

After the scalp deformation, we first move the hair roots by applying their barycentric coordinates to the deformed scalp vertices, and then project them onto the head surface $\mathcal{H}$ to eliminate the discretization error left. By leveraging the physics-based hyperelasticity energy, our relocation strategy enables a natural and smooth diffusion of the given hairline displacements over the whole scalp, while bringing minimal distortions to the original distribution of the hair roots.

\subsection{Shape Adaptation Tuning}
With the relocation of the hair roots ready, we still have to tune the shape adaptation in Section \ref{sec:hairstyle_transfer_method} for it. First, we change the right-hand side of the constraints in Equation \ref{eqn:root_constraint} to the relocated positions of hair roots. Second, the redistributed hair roots make it harder to balance the preservation of individual strand shapes and spatial relationships. Therefore, we introduce an adaptive weight for each non-root particle according to its curve-space position in the strand:
\begin{equation}
\label{eqn:per_particle_weight}
\gamma_{i} = 1 - e^{-\sigma(s_{i}/r_{i})^2},
\end{equation}
where $s_{i}$ is the arc length from particle ${i}$ to its hair root, $r_{i}$ is the traveling distance of the hair root during relocation, and $\sigma$ is a constant that we set to $0.2$. The closer a particle is to its hair root, the larger this weight is. Accordingly, we fine-tune the energy terms for the inter-strand and hair-body relationships:
\begin{equation}
\label{eqn:finetuned_energy}
\begin{split}
E_\mathsf{inter\text{-}strand} &= \sum_{i=0}^N \gamma_{i} \cdot \left\|\mathbf{L}_{i} - \mathbf{\bar{L}}_{i}\right\|_2^2 \\
E_\mathsf{hair\text{-}body} &= \sum_{i=0}^N \gamma_{i} \cdot \left\|\mathbf{p}_{i} - \mathbf{\hat{p}}_{i}\right\|_2^2.
\end{split}
\end{equation}
Harnessing such tuning to the optimization, we can produce a transfer result with a visually pleasant trade-off between the preservation of different shape factors, under a moderate hairline adjustment.

\section{Experimental Results}

\subsection{Implementation Details}
We build our system on a desktop PC with an Intel Core i9-13900K CPU and 128 GB memory. We implement the optimization module with the auto-differentiation tool Chumpy \cite{loper2014chumpy} and the OSQP solver \cite{osqp}. Moreover, we use ANN ~\cite{mount2010ann} to find the nearest neighbors and libigl \cite{jacobson2013libigl} to parameterize the head surface mesh. In Table \ref{tab:implementation}, we list all the hyperparameters for our method, and we set their values by using meters as the unit of model size.

\begin{table}[!h]
\centering
\renewcommand{\arraystretch}{1.2}
\caption{The hyper-parameters used in our algorithm.}
\begin{tabular}{l l r}
	\hline
	\textbf{Symbol} & \textbf{Description} & \textbf{Value} \\
	\hline\hline
	$\alpha$ & the weight of $E_\mathsf{inter\text{-}strand}$ in Eqn.(\ref{eqn:constrained_optimization}) & 3e3 \\
	$\beta$ & the weight of $E_\mathsf{hair\text{-}body}$ in Eqn.(\ref{eqn:constrained_optimization}) & 1e3 \\
	$k$ & the \#neighbors for $knn$ in Eqn.(\ref{eqn:laplacian_feature}) & 5 \\
	$\epsilon_\mathrm{c}$ & the clearance safeguard in Eqn.(\ref{eqn:penetration_constraint}) & 5e-4 \\
	$\epsilon_\mathrm{s}$ & the \revised{discrepancy} threshold in Eqn.(\ref{eqn:sharp_corners}) & 0.3 \\
\hline
\end{tabular}
\label{tab:implementation}
\end{table}

\subsection{Diversity and Consistency}
\begin{figure*}[!t]
\centering 
\includegraphics[width=1.0\textwidth]{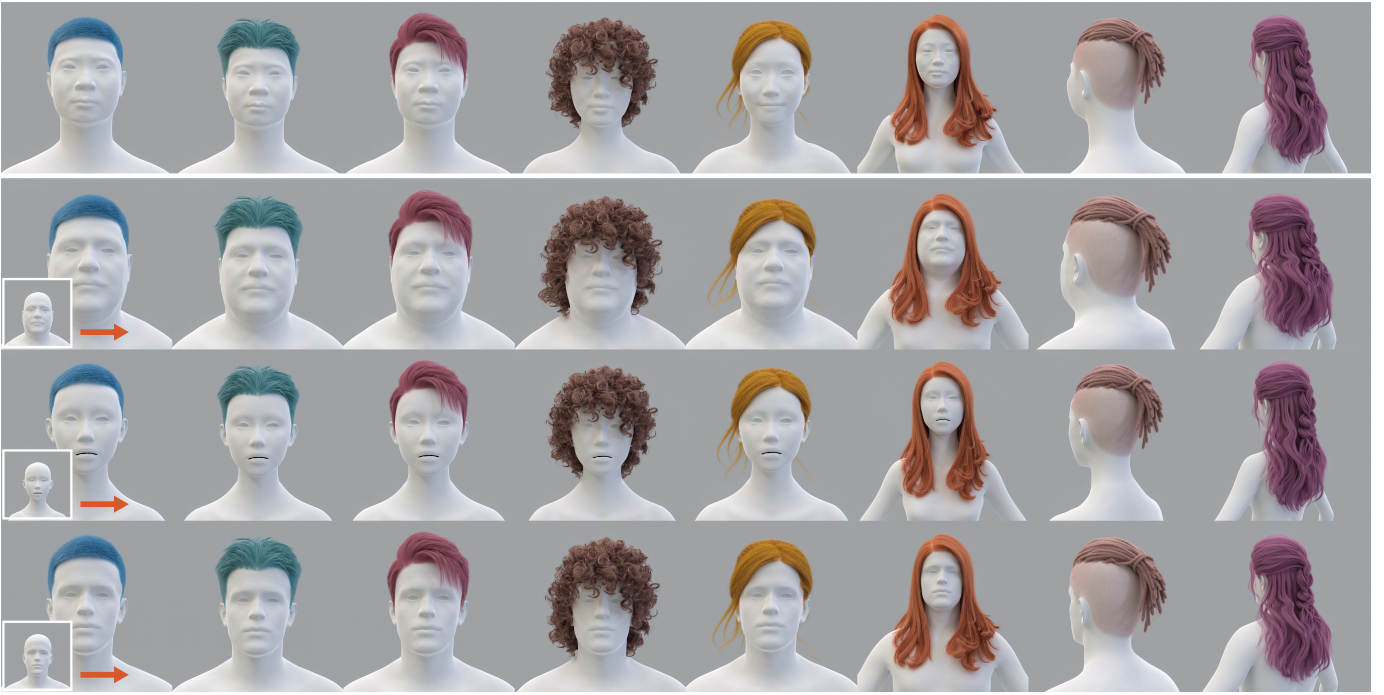}
\caption{\revised{Our method is capable of retargeting diverse hairstyles. The first row shows 8 source hairstyles, and the next rows show their adaptation variants by retargeting to 3 different target characters. See the supplemental for more results.}}
\label{fig:transfer_diversity}
\end{figure*}

Our method is capable of retargeting diverse 3D hairstyles to characters with distinct body shapes. We exemplify some results in Figure \ref{fig:transfer_diversity}, which contains $8$ hairstyles with different lengths, curliness, growth directions, split styles, and braid styles, and $3$ target characters with different ages, genders, and races. The transferred hairstyles faithfully preserve the shapes and spatial interactions of the sources.

\begin{figure}[!b]
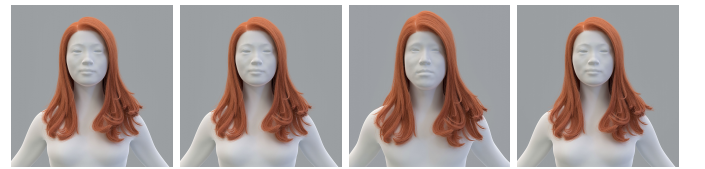

\centering
\foursubfig{1.0}
	{source A}{consistency/source.jpg}
	{A $\to$ A}{consistency/consistency_1_result.jpg}
 	{A $\to$ B}{consistency/consistency_2_mid.jpg}
  	{A $\to$ B $\to$ A}{consistency/consistency_2_result.jpg}
\caption{(a) The character A and its hairstyle. (b) \textit{reflexive test}: Retargeting the hairstyle from A to itself. (c) Retargeting the hairstyle from A to B. (d) \textit{cycle test}: Retargeting the hairstyle from A to B and then back to A.}
\label{fig:regression_test}
\end{figure}

We execute two regression experiments, i.e., the \textit{reflexive test} and \textit{cycle test}, to demonstrate the inherent consistency of our retargeting method. We show these regression results in Figure \ref{fig:regression_test}. Visually, they are almost indistinguishable from the original hairstyle. In Table \ref{tab:regression_error}, we report their corresponding per-particle distances and per-segment angle differences.

\begin{table}[!t]
\centering
\renewcommand{\arraystretch}{1.2}
\caption{Quantitative errors of the regression experiments.}
\begin{tabular}{l c c}
	\hline
 & \textbf{Mean Distance} & \textbf{Mean Angle}\\
	\hline\hline
\textbf{Regression 1 (reflexive)} &  $7.85\times10^{-8} \text{ m}$ & $3.75\times10^{-6} \text{ rad}$\\
\textbf{Regression 2 (cycle)} & $1.97\times10^{-3} \text{ m}$ & $4.64\times10^{-3} \text{ rad}$\\
\hline
\end{tabular}
\label{tab:regression_error}
\end{table}





\subsection{Ablation Study}
\label{sec:ablation_study}

\begin{figure*}[!t]
\centering 
\includegraphics[width=1.0\textwidth]{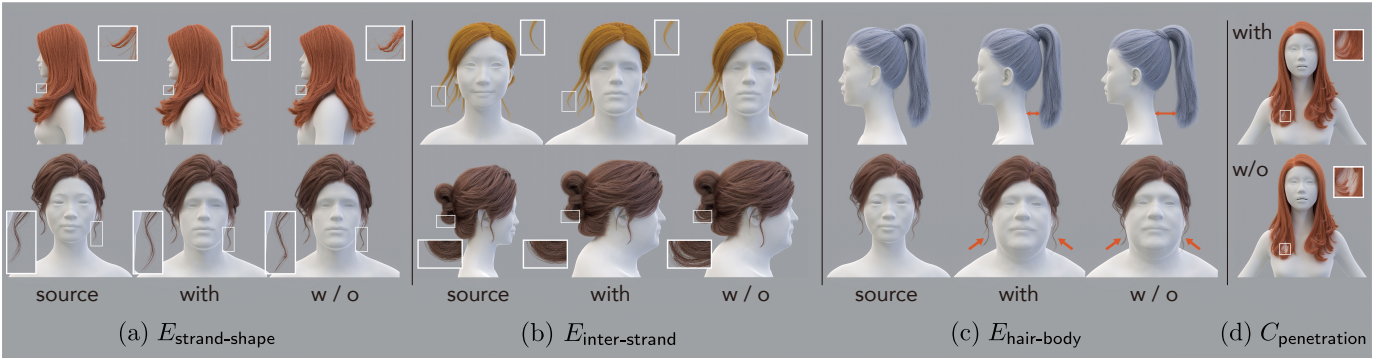}
\caption{Ablation study for the objectives and constraints used in the shape adaptation optimization (Equation \ref{eqn:constrained_optimization}).}
\label{fig:ablation_study}
\end{figure*}

\subsubsection{Constrained Optimization}
We conduct qualitative experiments to justify the necessity of individual energy terms and constraints in Equation \ref{eqn:constrained_optimization}.

Figure \ref{fig:ablation_study}a demonstrates the efficacy of the energy term $E_\mathsf{strand\text{-}shape}$ in Equation \ref{eqn:single_strand_shape}. Without it, the strand curvatures quickly distort during the optimization due to the preservation of other shape factors.

Besides the shapes of single strands, the spatial relationships between adjacent strands are significant visual clues to a hairstyle. Accordingly, Figure \ref{fig:ablation_study}b shows the efficacy of $E_\mathsf{inter\text{-}strand}$ in Equation \ref{eqn:strand_strand_relationship}. Without this term, the clustering and spread-out effects are easily damaged during the optimization.

Different characters have distinct shapes of head, face, neck, and shoulder. The spatial positionings of the strands relative to the body would quickly become inaccurate after retargeting if the energy term $E_\mathsf{hair\text{-}body}$ in Equation \ref{eqn:hair_body_relationship} is not incorporated during the optimization (see Figure \ref{fig:ablation_study}c).

Correct collision handling is crucial for successful retargeting. As shown in Figure \ref{fig:ablation_study}d, the interpenetration artifacts emerge soon if the hard constraint $C_\mathsf{penetration}$ in Equation \ref{eqn:penetration_constraint} is not incorporated.

\begin{figure}[!t]
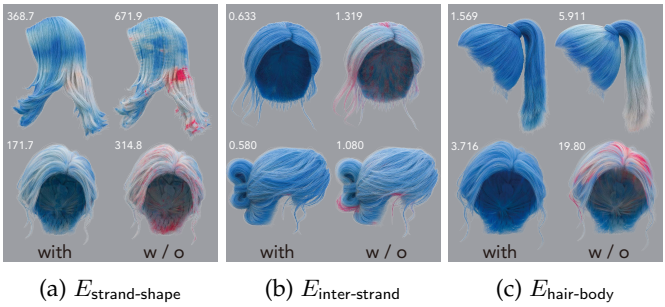

\centering
\threesubfig{1.0}
	{$E_\mathsf{strand\text{-}shape}$}{ablation_evaluation/hair_shape_compact_white_new.pdf}
 	{$E_\mathsf{inter\text{-}strand}$}{ablation_evaluation/hair_hair_compact_white_new.pdf}
	{$E_\mathsf{hair\text{-}body}$}{ablation_evaluation/hair_body_compact_white_new.pdf}
\caption{\revised{The objectives computed on results of Figure \ref{fig:ablation_study}.}}
\label{fig:ablation_evaluation_visualization}
\end{figure}

In contrast, with our complete formulation (Equation \ref{eqn:constrained_optimization}), all the energy terms and constraints mutually confine each other to achieve a balanced shape preservation. \revised{To provide a quantitative reference, Figure \ref{fig:ablation_evaluation_visualization} compares the per-particle and total objective values computed on results of Figure \ref{fig:ablation_study}.}

\subsubsection{Local Positioning Strategies}
\begin{figure}[!t]
\centering
\foursubfignoindex{1.0}
	{source}{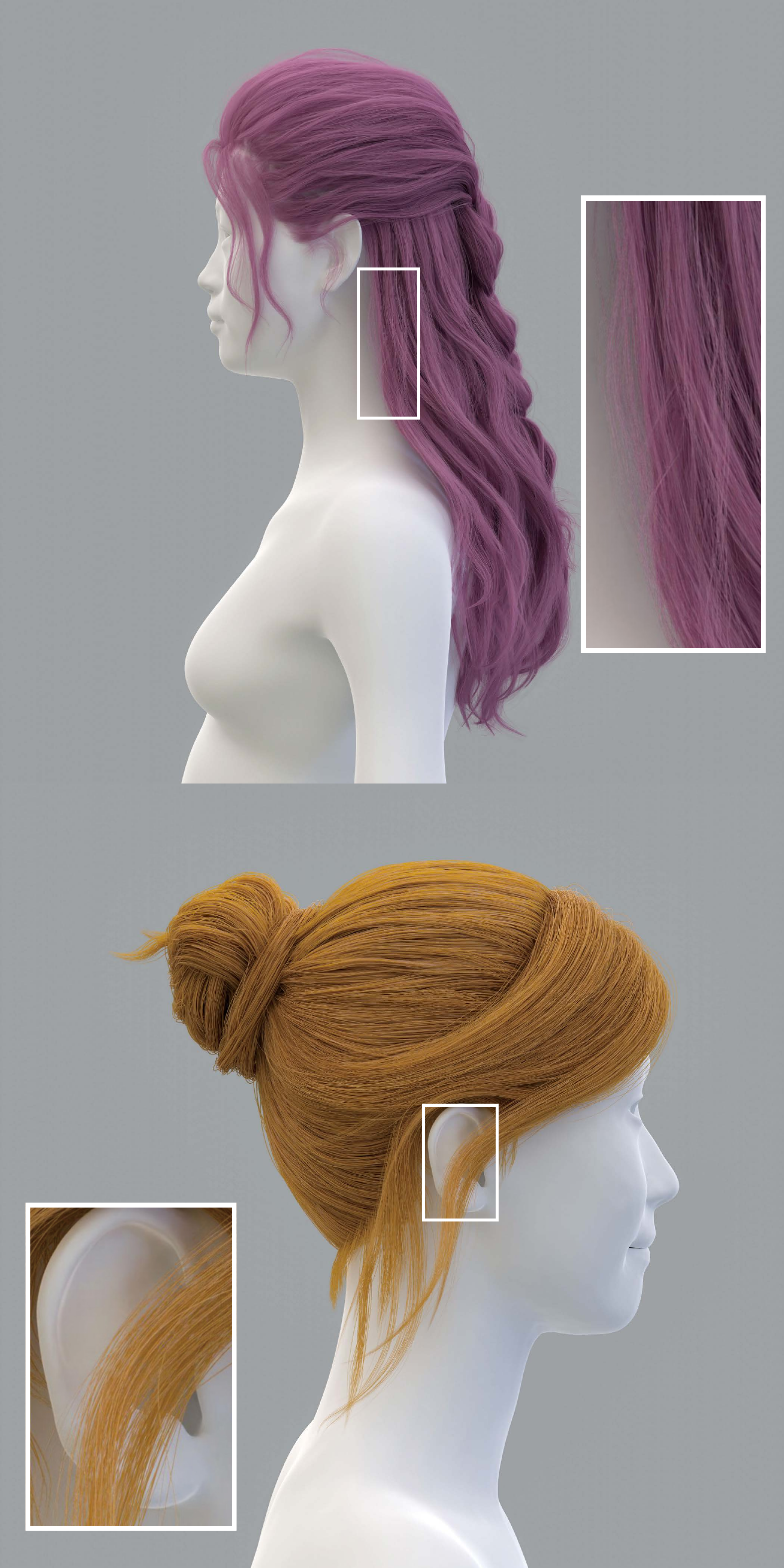}
	{modified \cite{brouet2012design}}{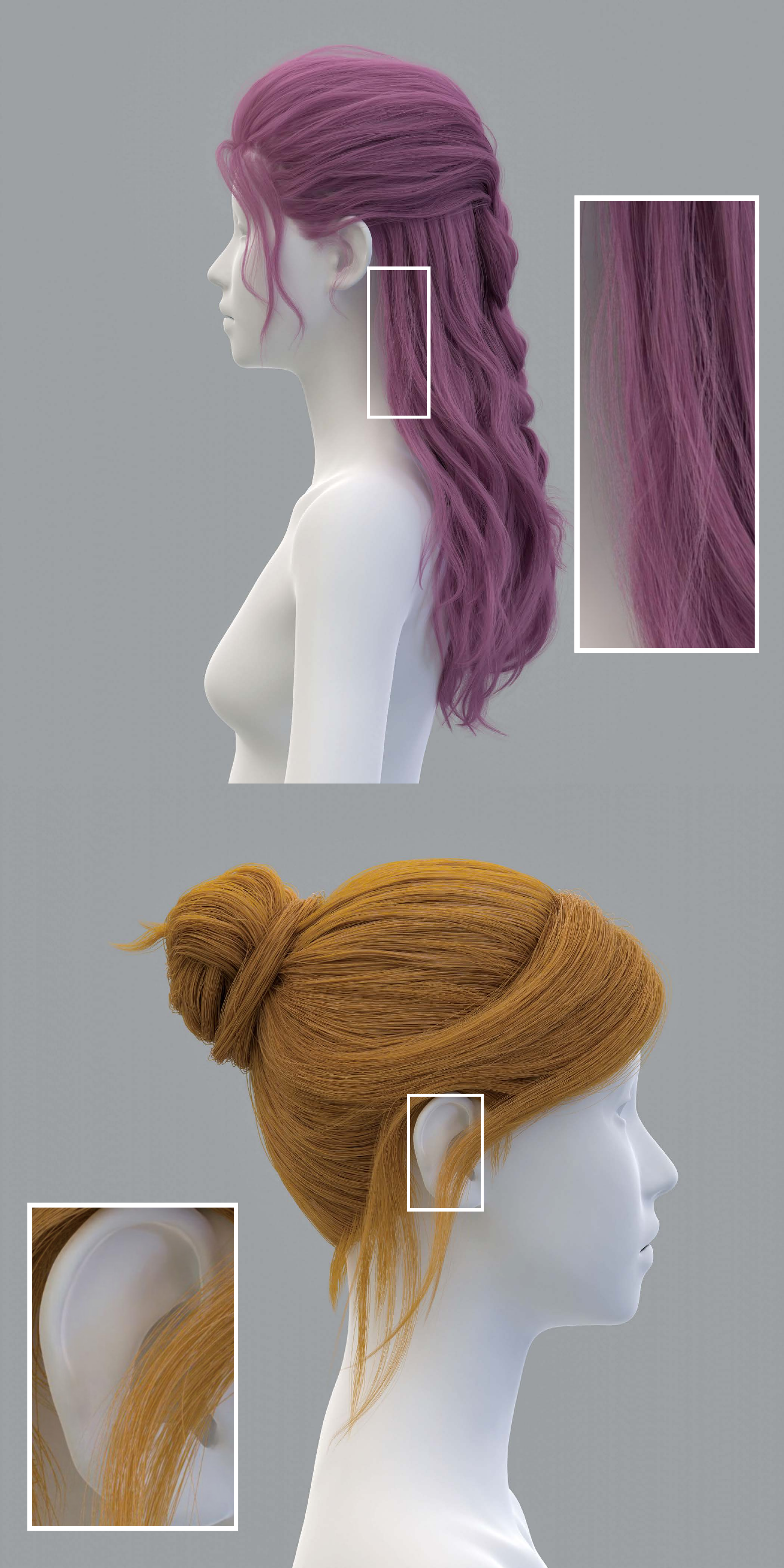}
	{\cite{al2013relationship}}{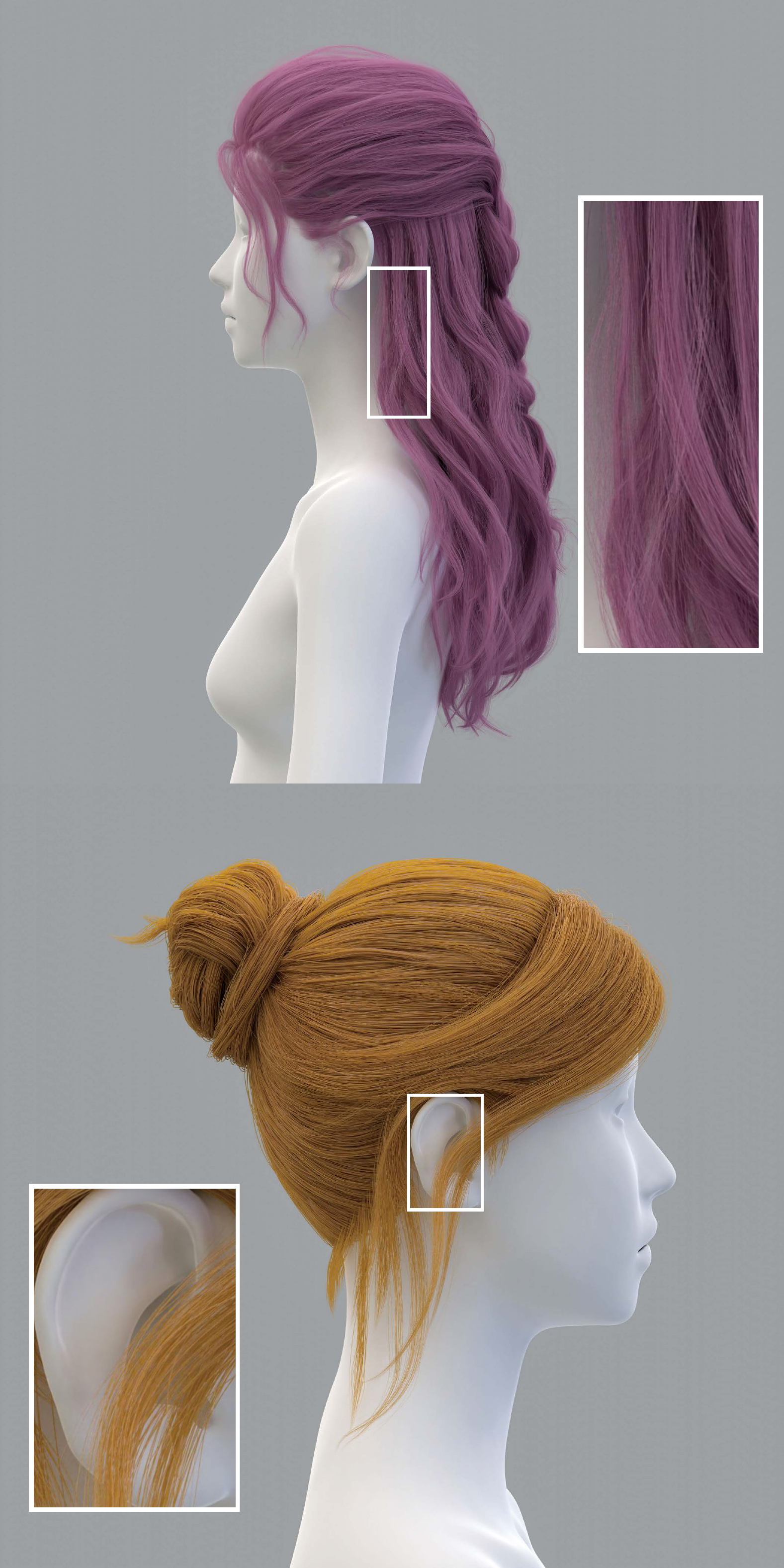}
	{\cite{meng2012flexible}}{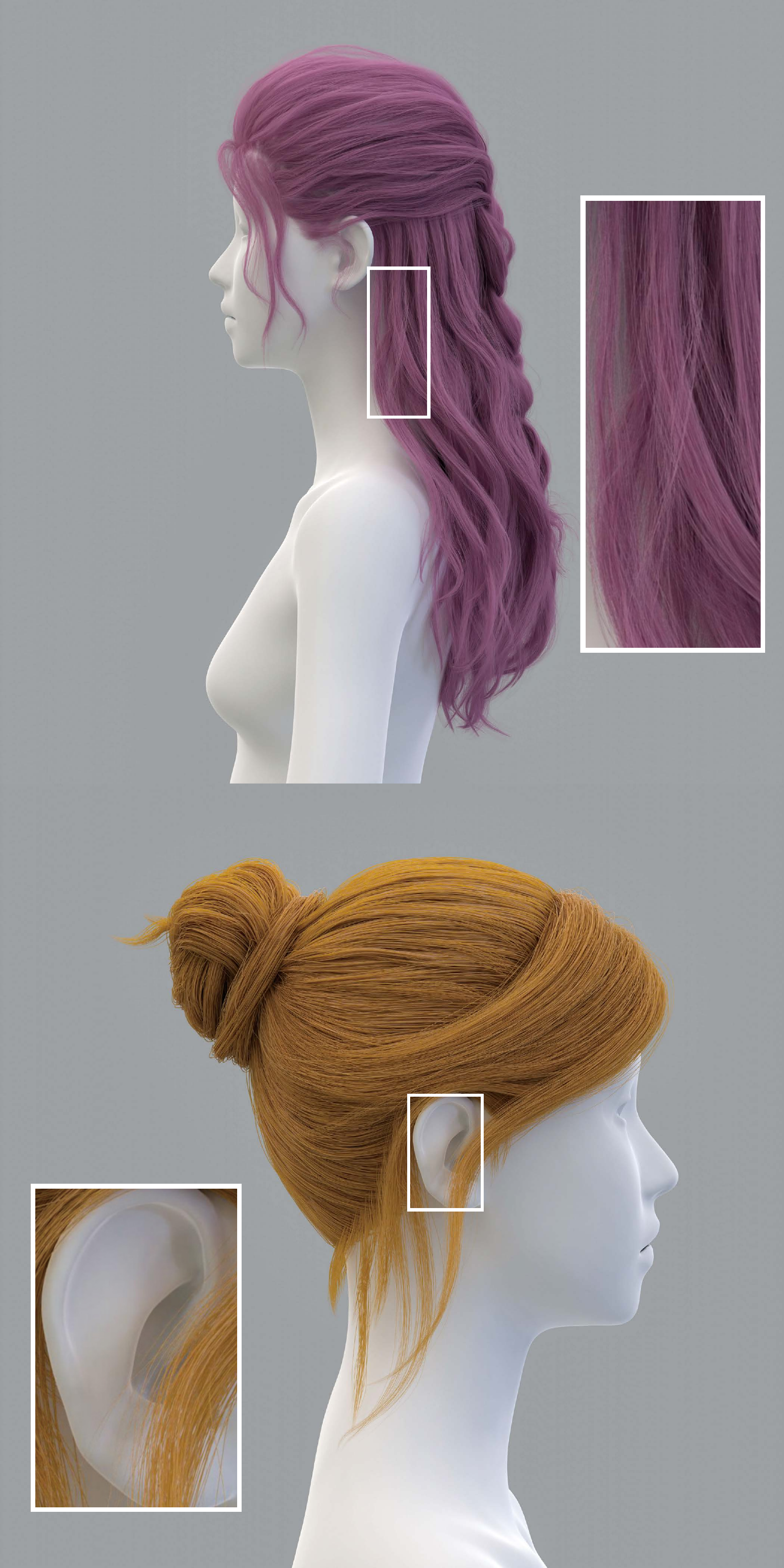}
\caption{Comparison of different local-positioning strategies.}
\label{fig:local_positioning_comparison}
\end{figure}

The local positioning of hair particles relative to the character body determines the quality of initial transfer (Section \ref{sec:initial_transfer}). In Figure \ref{fig:local_positioning_comparison}, we compare three local positioning strategies, i.e., \cite{meng2012flexible}, \cite{al2013relationship}, and our modified version of \cite{brouet2012design}. Specifically, \cite{meng2012flexible} finds the closest triangle on the body to each hair particle as its anchor, while \cite{al2013relationship} builds this correspondence by searching the coverage volume defined on each triangle, and the modified \cite{brouet2012design} selects the best bones to compute the positioning rays. We find that the modified \cite{brouet2012design} provides the most consistent local-positioning across all the hair particles, due to the fact that the body shapes usually differ much but the reference bones are rather stable.


\begin{figure}[!t]
\centering
\foursubfignoindex{1.0}
	{\cite{al2013relationship}}{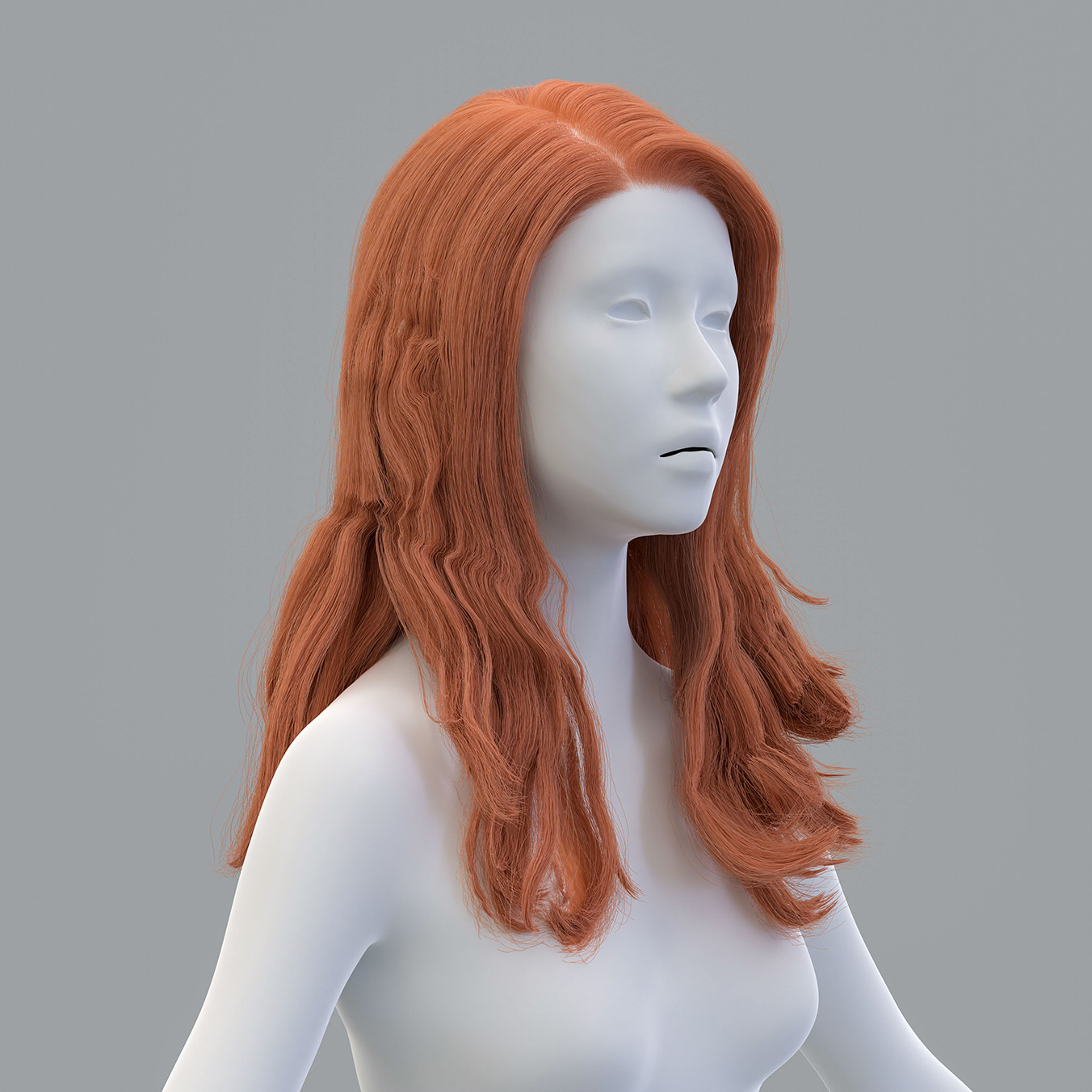}
	{\cite{meng2012flexible}}{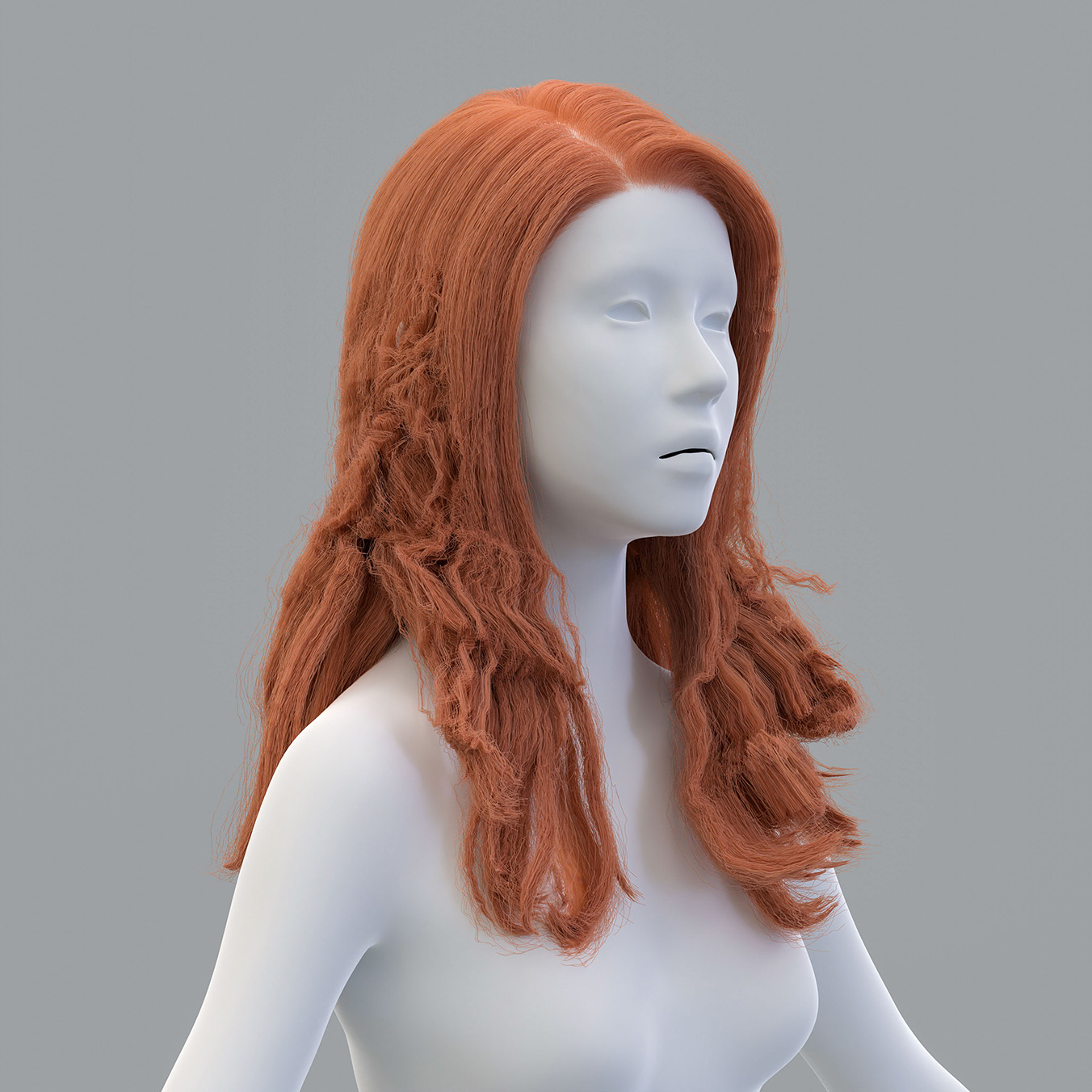}
	{\cite{brouet2012design}}{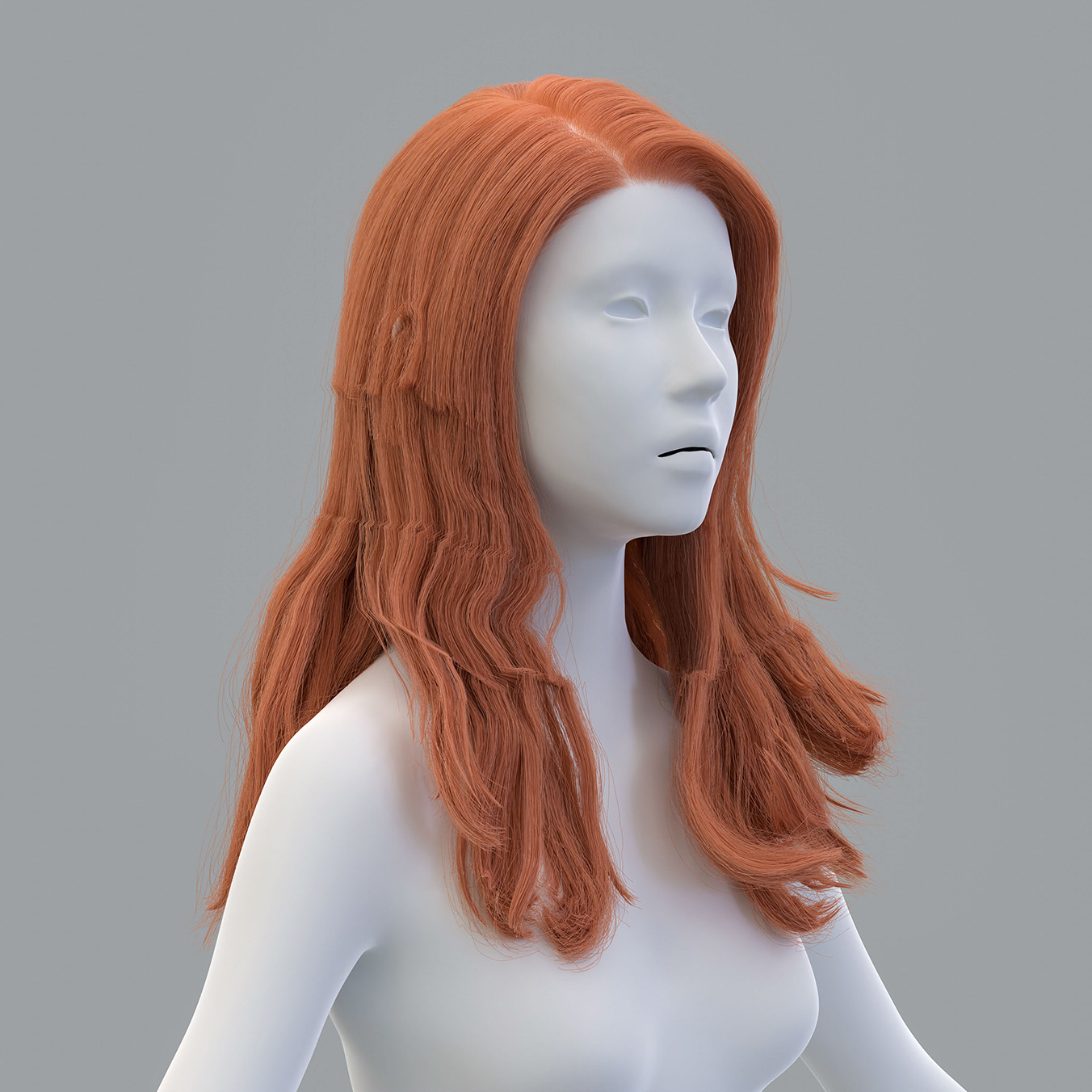}
	{\cite{brouet2012design} + smooth}{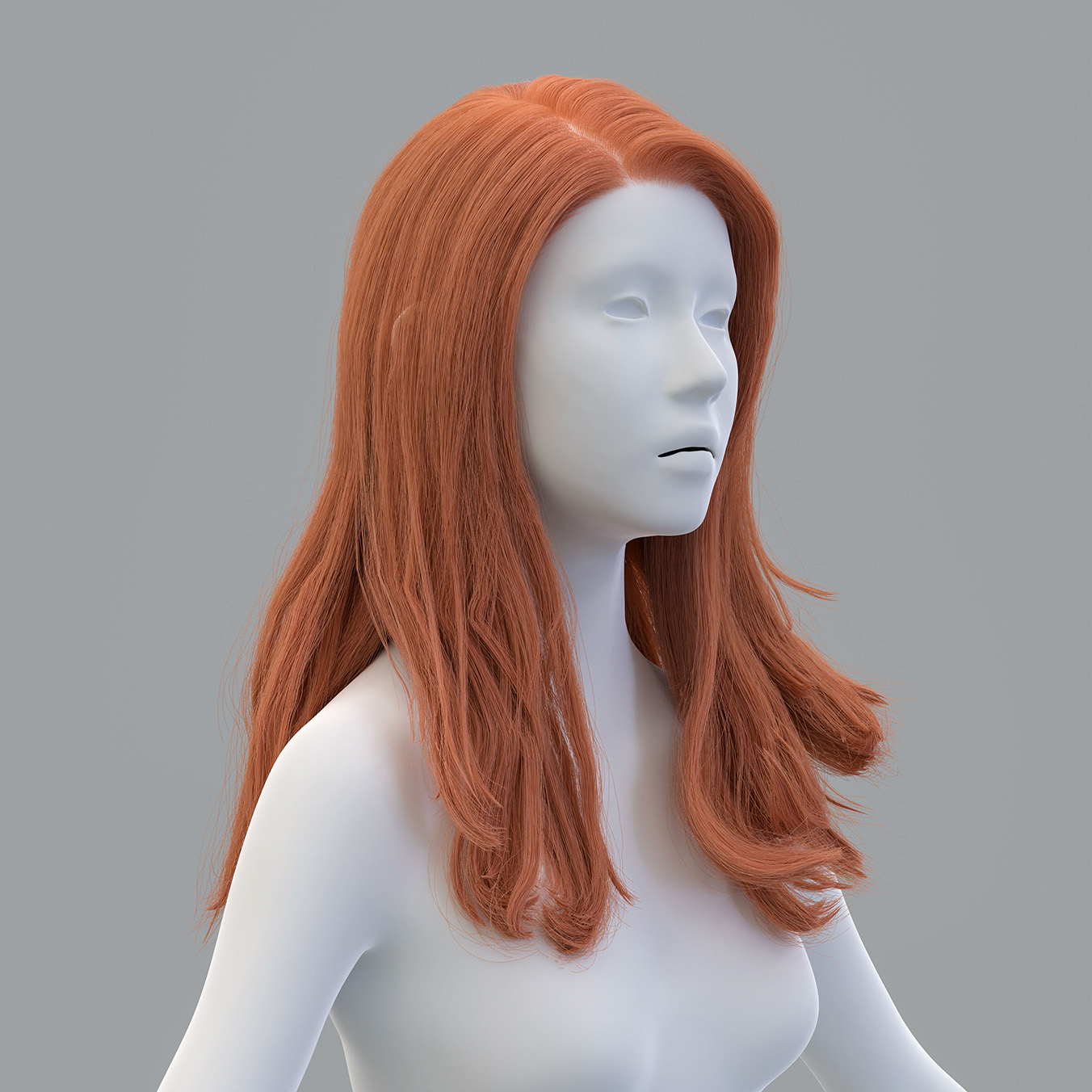}
\caption{Sharp turnings necessitates an adaptive smoothing.}
\label{fig:local_smoothing_comparison}
\end{figure}

Moreover, all the above local positioning strategies produce sharp turning points in the initial transfer result (Figure \ref{fig:local_smoothing_comparison}), whose smoothness is crucial for ensuring the convergence of the following optimizations. This necessitates an extra step for adaptive smoothing (Equation \ref{eqn:poisson_smooth}) right after recovering the local positionings on the target character.


\subsection{Multi-scale Transfer}
\begin{figure}[!t]
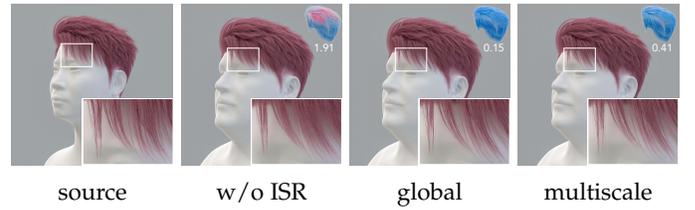

\centering
\foursubfignoindex{1.0}
	{source}{compare_multi_scale_and_global/source}
	{w/o ISR}{compare_multi_scale_and_global/wo_lap_v3}
	{global}{compare_multi_scale_and_global/global_v3}
	{multiscale}{compare_multi_scale_and_global/multi-scale_v3}
\caption{\revised{The adaptation without the inter-strand relationship (ISR) is over-smoothing. With ISR enabled, both the global and multi-scale solvers produce good fidelities. At the upper right, we show the per-particle and total value of $E_\mathsf{inter\text{-}strand}$. Though the multi-scale solver resorts to the guide hairs and optimizes a decoupled version of the inter-strand energy, a sufficient decrease is still obtained.}}
\label{fig:comp_multi_scale}
\end{figure}

The ablation study shows that preserving the inter-strands relationship is crucial for retargeting fidelity. However, the corresponding Laplacian energy requires solving a large linear system from the global coupling, which makes the whole optimization intractable for high-resolution hairstyles. For a typical curly hairstyle with $113$K strands and $2.9$M particles, the global solver consumes up to $86$GB of memory during optimization and requires over $1$ hour to complete the entire retargeting. Fortunately, our multi-scale solver introduced in Section \ref{sec:mss} consumes only $1.2$GB of memory and takes less than $2$ minutes to optimize, which is \textit{two orders of magnitude} faster. Figure \ref{fig:comp_multi_scale} shows that the retargeting results produced by the global and multi-scale solver do not have significant visual differences. See the supplemental for more results.



\subsection{Transfer with Hairline Edits}
\label{sec:hairline_deformation_experiment}
\begin{figure*}[!t]
\centering 
\includegraphics[width=1.0\textwidth]{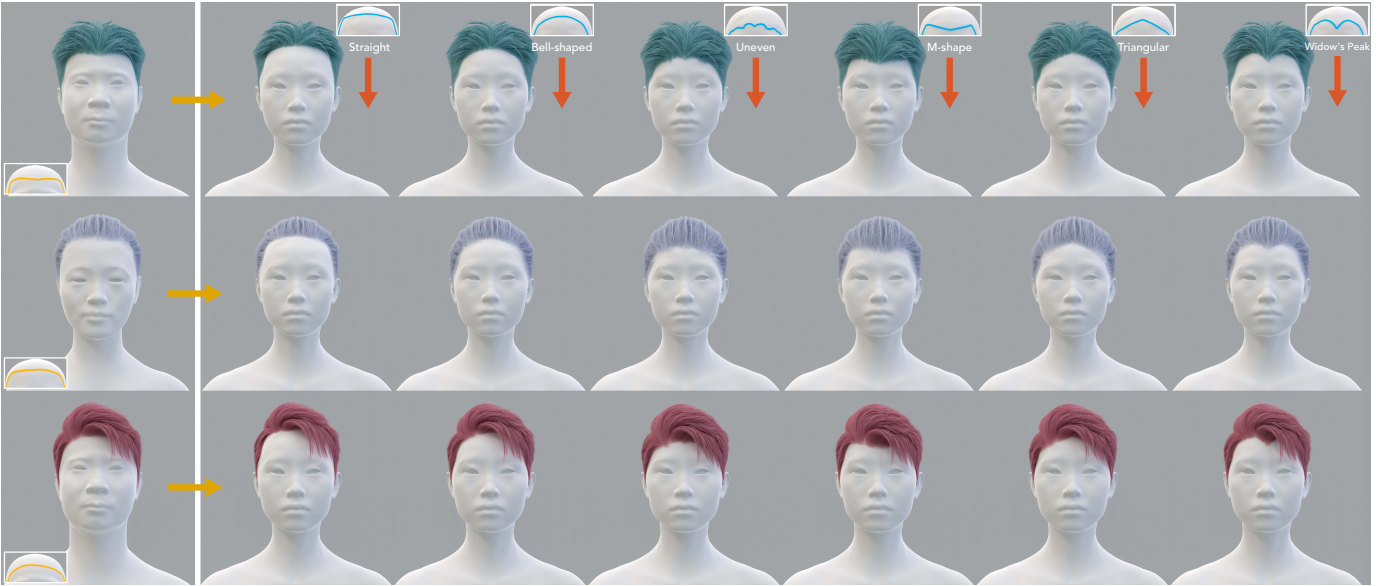}
\caption{Our method supports diverse hairline edits. The first column shows the source hairstyles. The following columns show their retargeting results to a new character with 6 common hairline types. See the supplemental for more results.}
\label{fig:hairline_change_diversity}
\end{figure*}

\begin{figure}[!t]
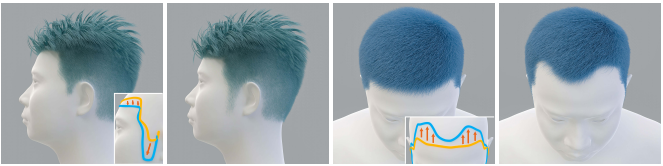

\centering
\foursubfignosubcap
	{hairline_edit/side_source.pdf}
	{hairline_edit/side_target.pdf}
	{hairline_edit/front_source.pdf}
	{hairline_edit/front_target.pdf}
\caption{The sideburns (left) and baldness (right) edits.}
\label{fig:hairline_edit_more}
\end{figure}

Our method supports diverse hairline edits. Figure \ref{fig:hairline_change_diversity} shows the retargeting results of 3 distinct hairstyles for 6 common hairline types, and Figure \ref{fig:hairline_edit_more} shows more edits.

\begin{figure}[!t]
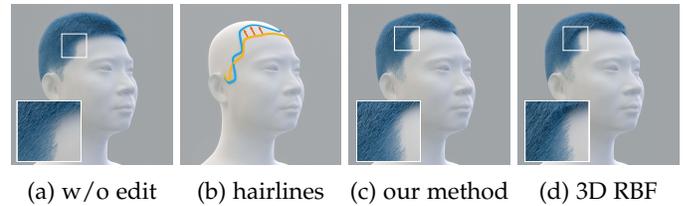

\centering
\foursubfig{1.0}
	{w/o edit}{hairline_deformation/rbf_3_compare/wo_hairline_deformation.pdf}
 	{hairlines}{hairline_deformation/rbf_3_compare/two_hairline.pdf}
	{our method}{hairline_deformation/rbf_3_compare/neo_hookean_lscm.pdf}
 	{3D RBF}{hairline_deformation/rbf_3_compare/rbf_3.pdf}
\caption{(a) The transfer result without hairline edits. (b) The {\color{cc3}source}/{\color{cc1}target} hairlines. (c-d) Hairline edits with our embedded membrane and 3D RBF interpolation, respectively.}
\label{fig:comp_3d_rbf}
\end{figure}

\begin{table}[!t]
\centering
\renewcommand{\arraystretch}{1.2}
\caption{Density changes of the hair roots. For 10 hairstyles under 2 hairline edits (S for straight, W for widow’s peak), we compute the $L_1$ and $L_{\infty}$ norms of the density change vectors and report the \textit{min}, \textit{max}, and \textit{avg} values across all the combinations. Compared with the alternatives (Figure \ref{fig:hc_diff_param_diff_method}), our method produces much smaller distortions consistently. LSCM \cite{levy2023least} is used for parameterization. See the supplemental for per-hairstyle data.}
\resizebox{\columnwidth}{!}{%
\begin{tabular}{@{\extracolsep{3pt}} l c c c c c c}
\hline
& \multicolumn{3}{c}{\textbf{$\|\cdot\|_1$ of density change}} & \multicolumn{3}{c}{\textbf{$\|\cdot\|_\infty$ of density change}}\\
\cline{2-4} \cline{5-7}
& \textbf{Our} & \textbf{Harmonic} & \textbf{RBF(2D)} & \textbf{Our} & \textbf{Harmonic} & \textbf{RBF(2D)} \\
\hline\hline
\textbf{min(S)} & \textbf{0.015} & 0.034 & 0.041 & \textbf{0.279} & 0.369 & 0.322 \\
\textbf{max(S)} & \textbf{0.048} & 0.122 & 0.152 & \textbf{1.023} & 1.706 & 2.886 \\
\textbf{avg.(S)} & \textbf{0.035} & 0.073 & 0.081 & \textbf{0.457} & 0.672 & 0.719 \\
\textbf{min(W)} & \textbf{0.019} & 0.034 & 0.047 & \textbf{0.280} & 0.435 & 0.372 \\
\textbf{max(W)}  & \textbf{0.052} & 0.125 & 0.126 & \textbf{0.991} & 1.706 & 2.742 \\
\textbf{avg.(W)} & \textbf{0.035} & 0.071 & 0.080 & \textbf{0.447} & 0.816 & 0.755 \\
\hline
\end{tabular}

}
\label{tab:density_change}
\end{table}

The hair roots relocation step introduced in Section \ref{sec:hair_roots_relocation} is the key to retargeting with hairline edits. We thus compare our embedded membrane formulation in Equation \ref{eqn:membrane_on_head} with several alternative models to show its superiority.

In Figure \ref{fig:comp_3d_rbf}, we compare to an RBF-based 3D interpolation method, choosing the thin plat spline, i.e., TPS, as the radial basis function. Specifically, we sample a sparse set of vertices inside the scalp to fix, i.e., zero displacements. Then, we combine them with the hairline vertices as the seeds and use their prescribed 3D displacements as the known data to fit an 3D displacement field based on RBF. Finally, we apply this 3D displacement field to all the free vertices of the scalp and reproject the displaced vertices onto the head to obtain the relocated hair roots. Figure \ref{fig:comp_3d_rbf}d shows that this method produces unnatural compression and distortion of the hair distribution.

\begin{figure*}[!t]
\centering 
\includegraphics[width=1.0\textwidth]{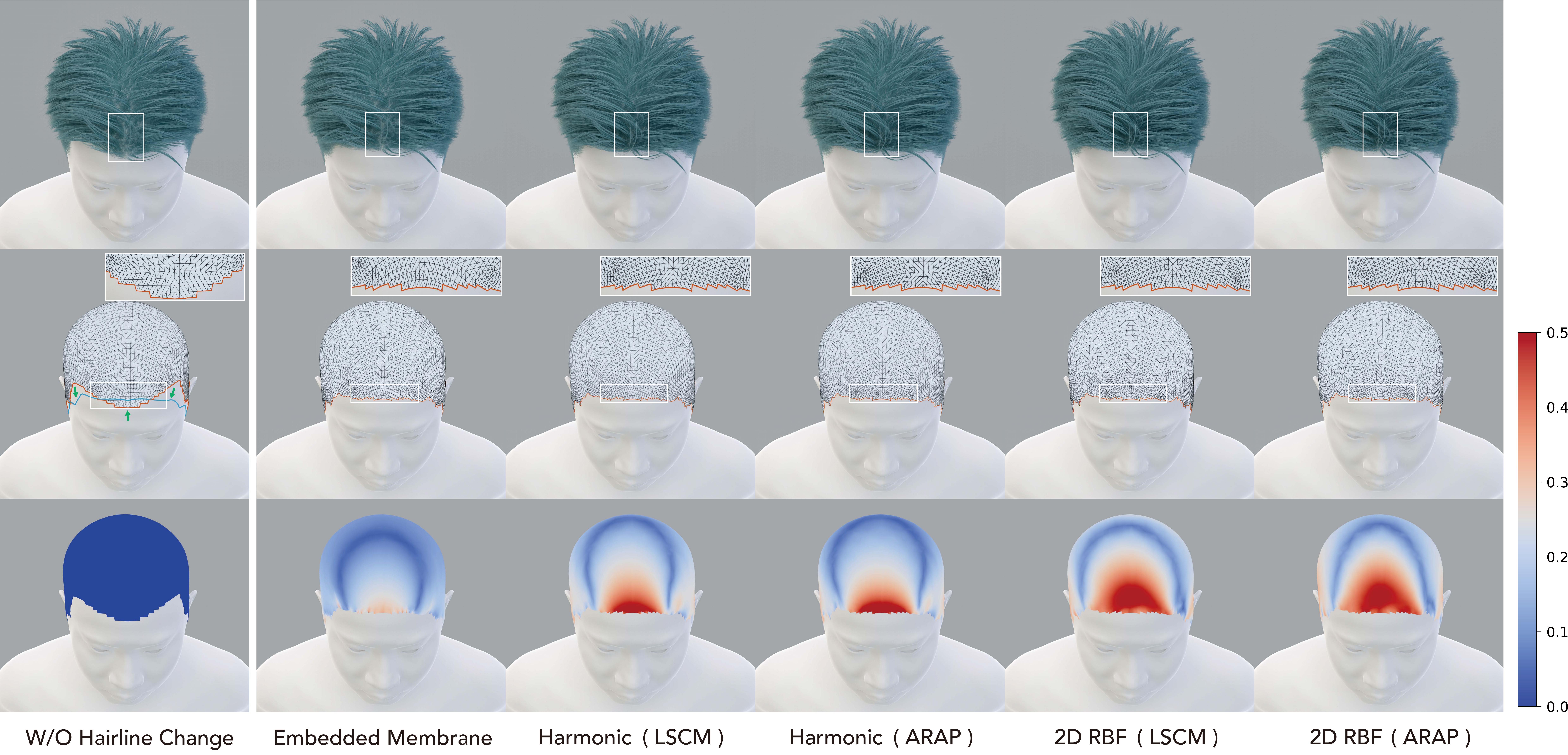}
\caption{From up to bottom: the produced 3D hairstyles, the deformed scalp mesh, and the hair-root density changes. The first column is the ground truth, i.e., retargeting without hairline edits. The following columns show the retargeting results with the same hairline edit (see row 2, column 1), using different hair-root relocation methods. The name in the parenthesis represents the surface parameterization method used. Compared with our embedded membrane method, the alternatives produce much more compressions and distortions in the scalp (see both the meshes and the density errors).}
\label{fig:hc_diff_param_diff_method}
\end{figure*}

The reprojection mentioned above is necessary because the interpolated vertex displacement is free in 3D, while the head surface is intrinsically 2D. To avoid such reprojection, we instead build a 2D displacement field in the parameterization space of the head. Specifically, we experiment with two different methods. One is to use an RBF interpolation as above but in 2D. The other is to solve a harmonic function on the parameterized scalp mesh with the same boundary conditions as our embedded membrane formulation. Both methods produce a 2D displacement field to move the scalp vertices in the parameterization space and recover their 3D positions. Since the hair-root relocation effects of these two methods depend on the parameterization, we compare two choices in Figure \ref{fig:hc_diff_param_diff_method}, i.e., LSCM\cite{levy2023least} and ARAP\cite{liu2008local}.

The uniformity of the scalp deformation determines the quality of the hair-root relocation. Figure \ref{fig:hc_diff_param_diff_method} compares the scalp mesh deformation generated by our embedded membrane and all the alternatives. In this figure, we also show the density change of the hair roots over the scalp region. 
Specifically, for each triangle, we define the average density as the number of hair roots in this triangle divided by its area. We compute per-triangle density values on the initial and deformed mesh, respectively, and stack all their relative changes into a vector. Table \ref{tab:density_change} sums up the $L_1$ and $L_{\infty}$ norms of the density change vectors across 10 hairstyles under 2 different hairline edits. The smaller the relative change, the less distortion the deformation introduces, and the smoother the hair-root relocation is.

\begin{figure}[!t]
\centering
\foursubfignoindex{1.0}
	{ARAP}{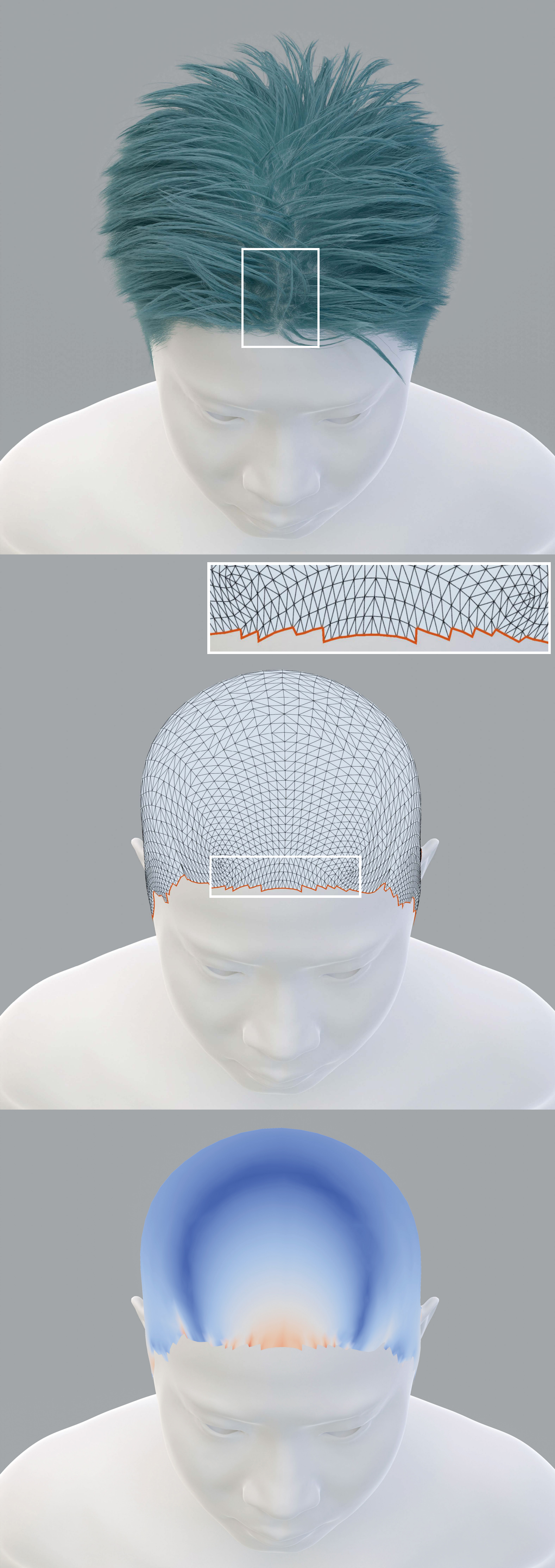}
 	{Harmonic}{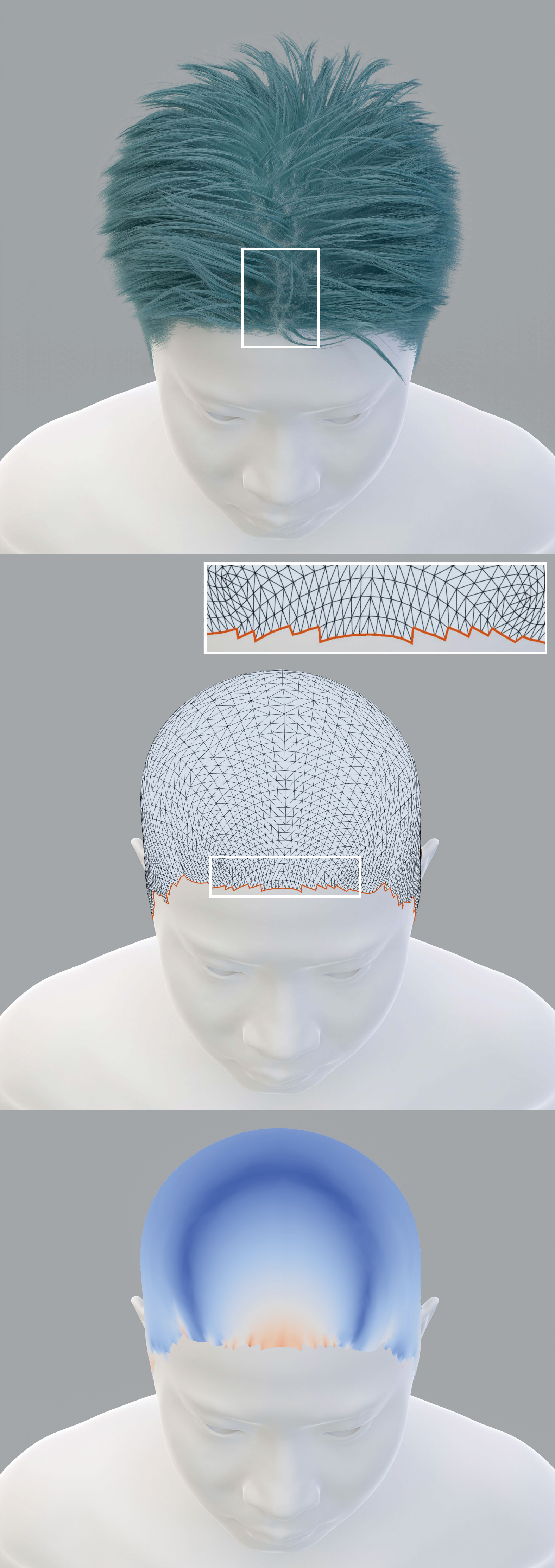}
	{LSCM}{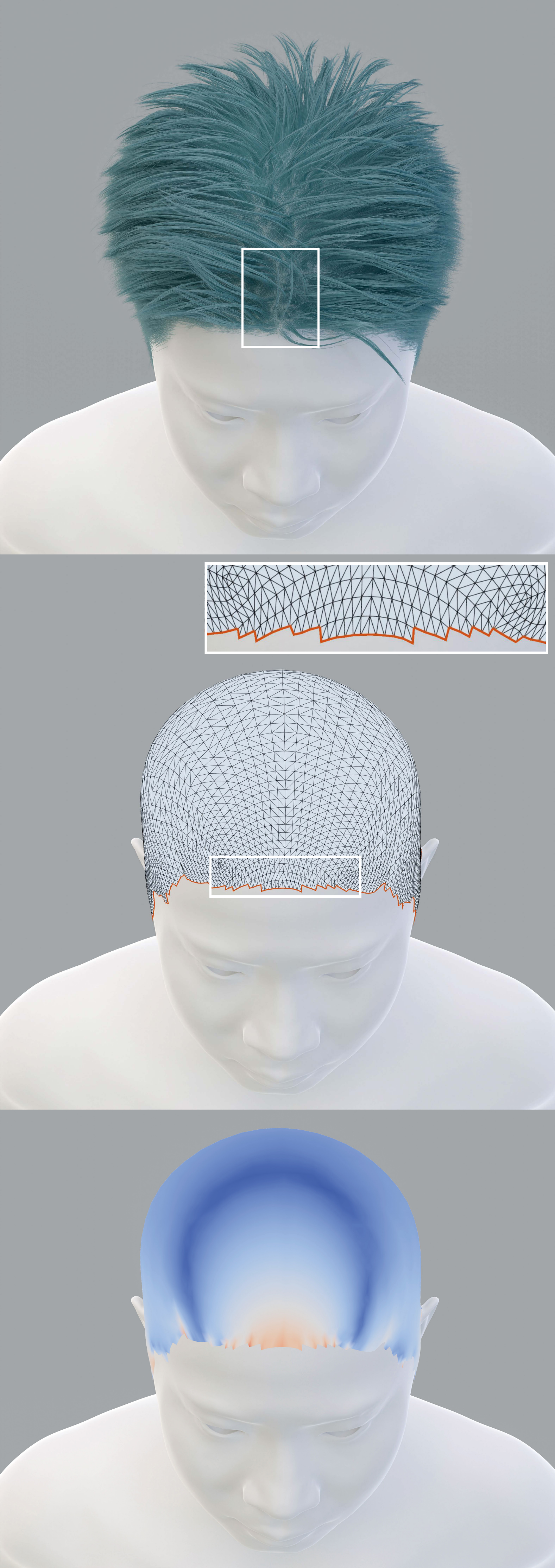}
 	{Method in \cite{wang2009example} \phantom{xx}}{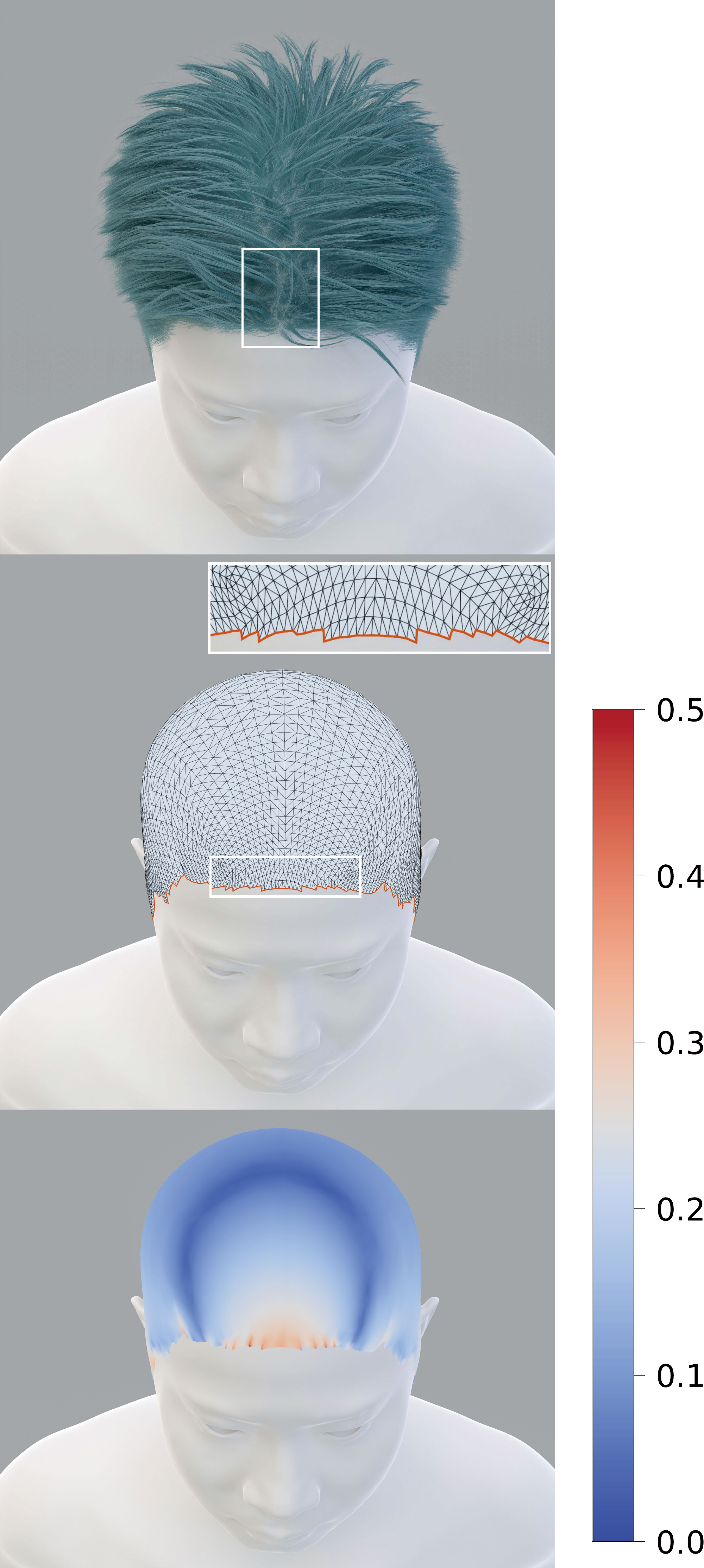}
\caption{Retargeting results under the same hairline edit by using our embedded membrane with four different parameterizations, i.e., ARAP\cite{liu2008local}, Harmonic map \cite{eck1995multiresolution}, LSCM \cite{levy2023least}, and the one introduced in \cite{wang2009example}.}
\label{fig:hc_diff_param_neo_hookean}
\end{figure}

Though our embedded membrane also requires a surface parameterization to optimize the scalp vertex, the deformation energy itself is defined in the 3D space. Therefore, our method always converges to similar results no matter which parameterization algorithm is used. Figure \ref{fig:hc_diff_param_neo_hookean} compares the scalp deformation with four popular parameterizations, and the results are consistent.


\subsection{Performance}
In practice, given the source character and the 3D hairstyle customized for it, we first extract the necessary information in advance for subsequent retargeting tasks. Such preprocessing depends on these source models only and is agnostic of any other characters. Specifically, we compute the local references and coordinates for hair particle positioning with respect to the body, the guide hairs for multi-scale solving, and the knn sets for hair Laplacian. Harnessing the extracted information, we obtain an \textit{adaptable} hairstyle ready for retargeting. At runtime, we adapt this hairstyle to an arbitrary target character, with an optional user edit of the hairline, by solving the constrained optimizations (Equation \ref{eqn:constrained_optimization}, \ref{eqn:membrane_on_head}).

Table \ref{tab:performance} shows the model complexities of all the experimented 3D hairstyles and their transfer time. The QP solving dominates the runtime computation. Although the most complex hairstyle contains more than $110K$ strands and $8M$ particles, our method produces a high-fidelity transfer result in $5$ minutes at runtime, ensuring its practical use in digital human applications.

\begin{table*}[!t]
\centering
\renewcommand{\arraystretch}{1.2}
\caption{\textbf{Runtime Statistics.} Columns from left to right: the model name, the number of hair strands/particles, the number of guide strands, the preprocessing time, the initial transfer time, the hair-root relocation time, the multi-scale solving time, and the total runtime (in seconds). We also list the full global solving time and the speedup of the multi-scale solving.}
\resizebox{\textwidth}{!}{
\begin{tabular}{l r r r r r r r r r}
	\hline
 	& & & & \multicolumn{4}{c}{\textbf{Runtime}} &\\
	\cline{5-8}
	\addlinespace[3pt]
	\textbf{Hairstyle} & \textbf{\#Strands / \#Particles} & \textbf{\#Guides} & \textbf{Preprocess} & \makecell{\textbf{Initial} \\ \textbf{Transfer}} & \makecell{\textbf{Hair-root} \\ \textbf{Relocation}} &\makecell{\textbf{Multi-scale} \\ \textbf{Solving}} &\textbf{Total} & \makecell{\textbf{Full} \\ \textbf{Solving}} & \textbf{Speedup}\\
	\hline\hline
	Short hair (Fig.\ref{fig:transfer_diversity}, col 1) & 65,819 / 460,376 & 302 & 6.66s & 6.35s & N/A & 14.20s & 20.55s & 440.58s & 31$\times$\\
	Dreadlocks(Fig.\ref{fig:transfer_diversity}, col 7) & 101,763 / 4,986,476 & 309 & 74.37s & 21.41s & N/A & 113.73s & 135.14s & 6,960.05s & 61$\times$\\
	Spiky (Fig.\ref{fig:transfer_diversity}, col 2) & 109,097 / 2,218,745 & 318 & 30.30s & 11.63s & N/A & 52.12s & 63.75s & 3,061.56s & 59$\times$\\
	Side fringe (Fig.\ref{fig:transfer_diversity}, col 3) & 112,962 / 2,961,992 & 302 & 44.0s & 14.67s & N/A & 73.65s & 88.32s & 3,940.99s & 54$\times$\\
	Ponytail (Fig.\ref{fig:ablation_study}, row 1, col 7) & 85,259 /  2,837,372 & 297 & 61.98s & 12.90s & N/A & 51.59s &  64.49s &  5,726.29s & 111$\times$\\
	Long hair 1 (Fig.\ref{fig:transfer_diversity}, col 6) & 49,991 / 7,229,645 & 312 & 165.66s & 38.41s & 21.23s &  148.70s & 208.34s & 13,810.93s& 93$\times$\\
	Long hair 2 (Fig.\ref{fig:transfer_diversity}, col 8) & 96,089 / 8,473,200 & 316 &  105.52s & 34.47s & 25.92s & 207.79s & 268.18s & 8,021.95s&39$\times$\\
	Updo 1 (Fig.\ref{fig:overview}) & 86,847 /  5,542,057 & 308 & 122.62s & 21.30s &26.71s &177.80s & 225.81s & 10,337.34s&58$\times$\\
    Curly (Fig.\ref{fig:transfer_diversity}, col 4) &  117,364 / 4,456,984 & 316 &  121.66s &15.24s&22.67s&130.31s & 168.22s & 9,643.62s & 57$\times$\\
	Updo 2 (Fig.\ref{fig:transfer_diversity}, col 5) & 111,042 / 6,769,042 & 302 & 167.06s & 26.27s & 23.34s& 205.53s &255.14s & 18,245.06s& 89$\times$\\
	Med-Len (Fig.\ref{fig:teaser}) &84,268 / 6,207,432 & 329 & 113.78s & 23.65s & 20.03s& 120.52s & 164.20s & 19,890.22s & 165$\times$\\
\hline
\end{tabular}
}
\label{tab:performance}
\end{table*}

\begin{figure}[!b]
\centering
\foursubfignosubcap
	{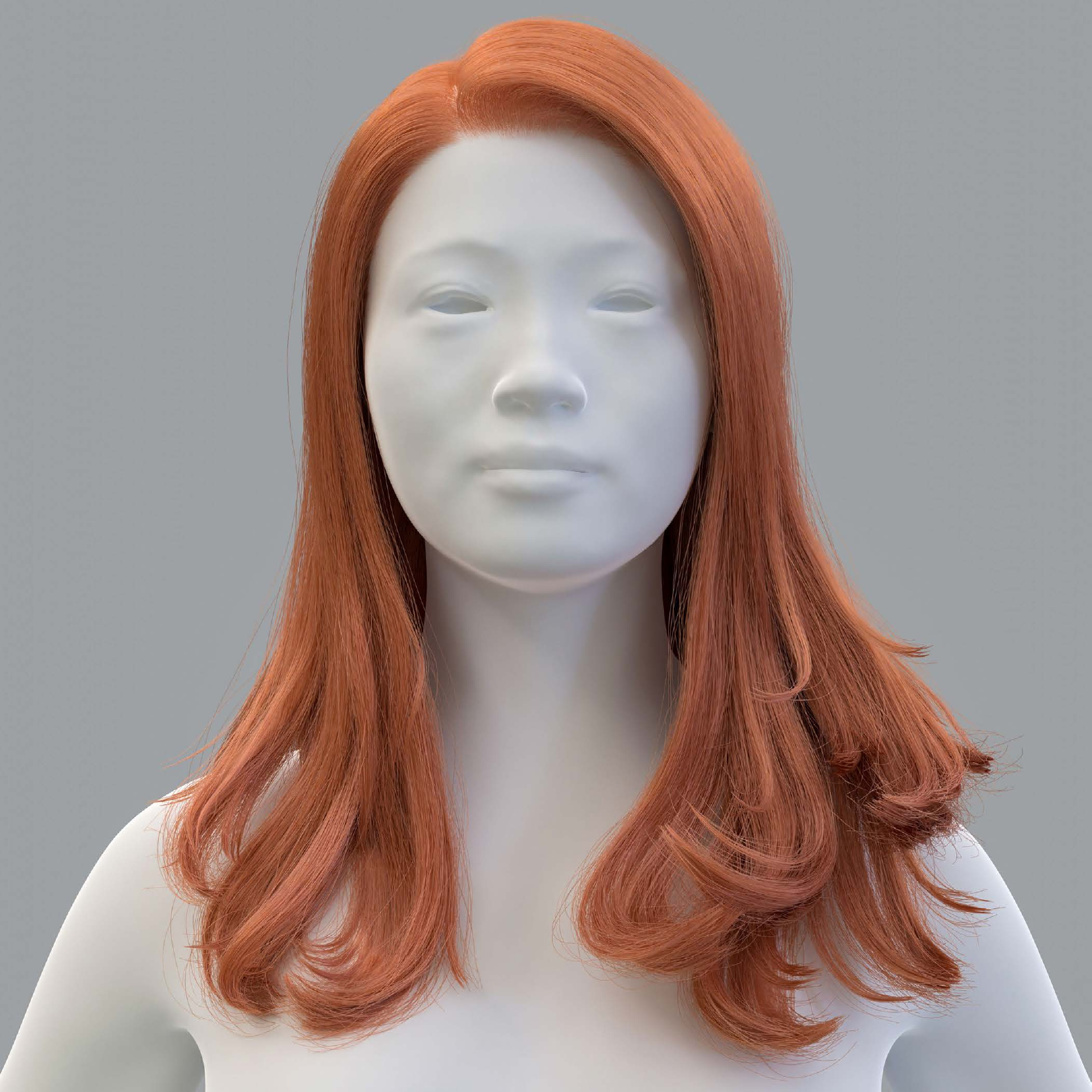}
	{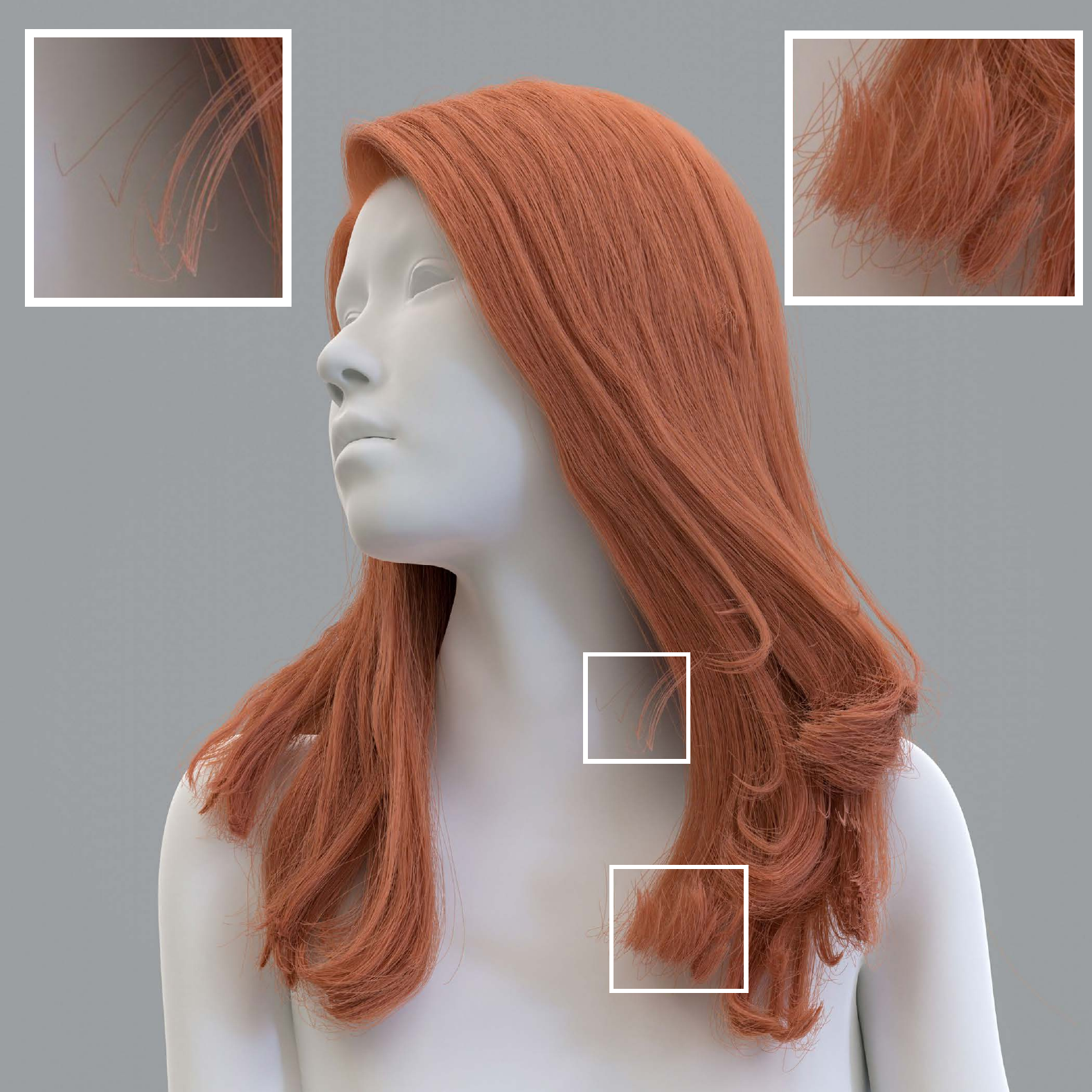}
	{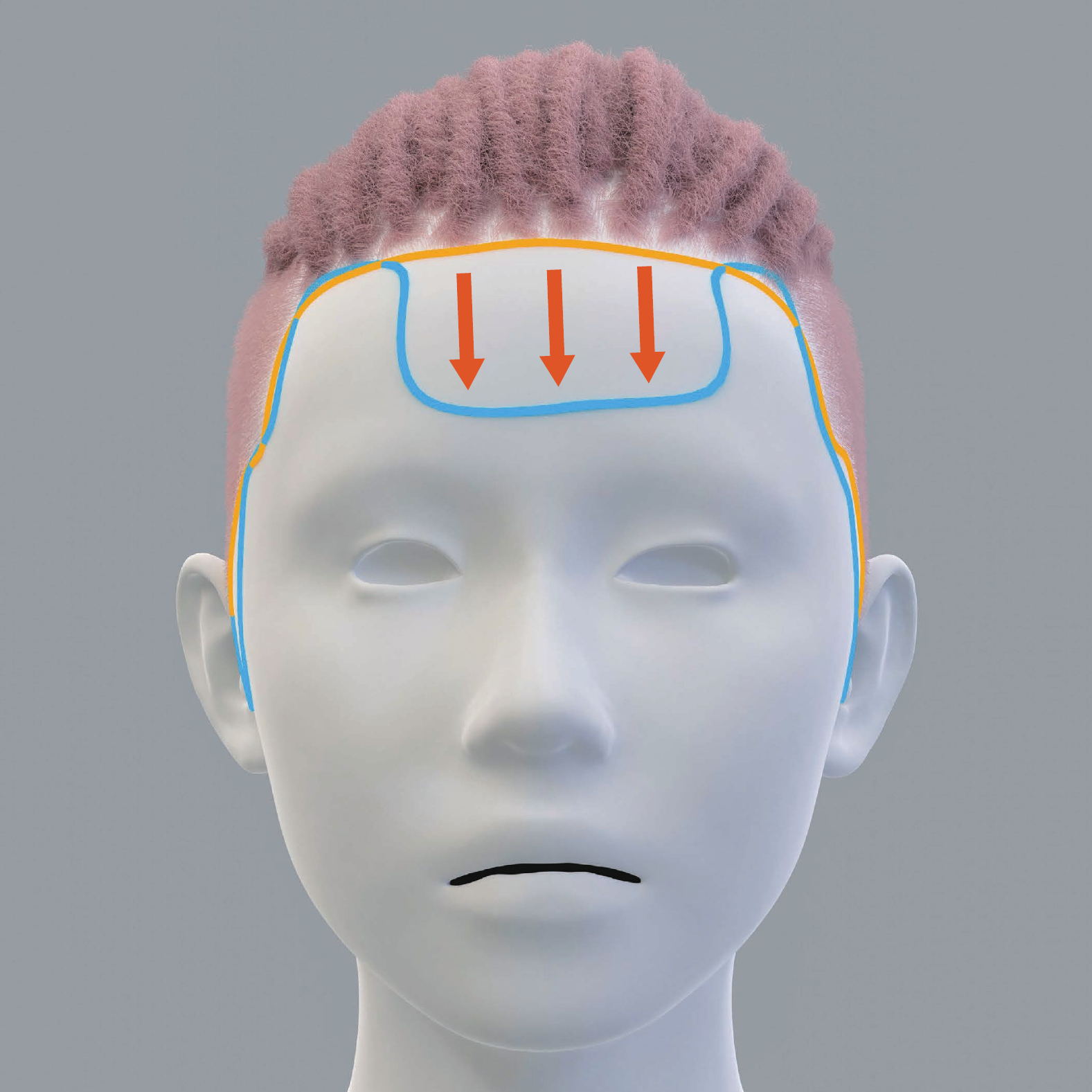}
	{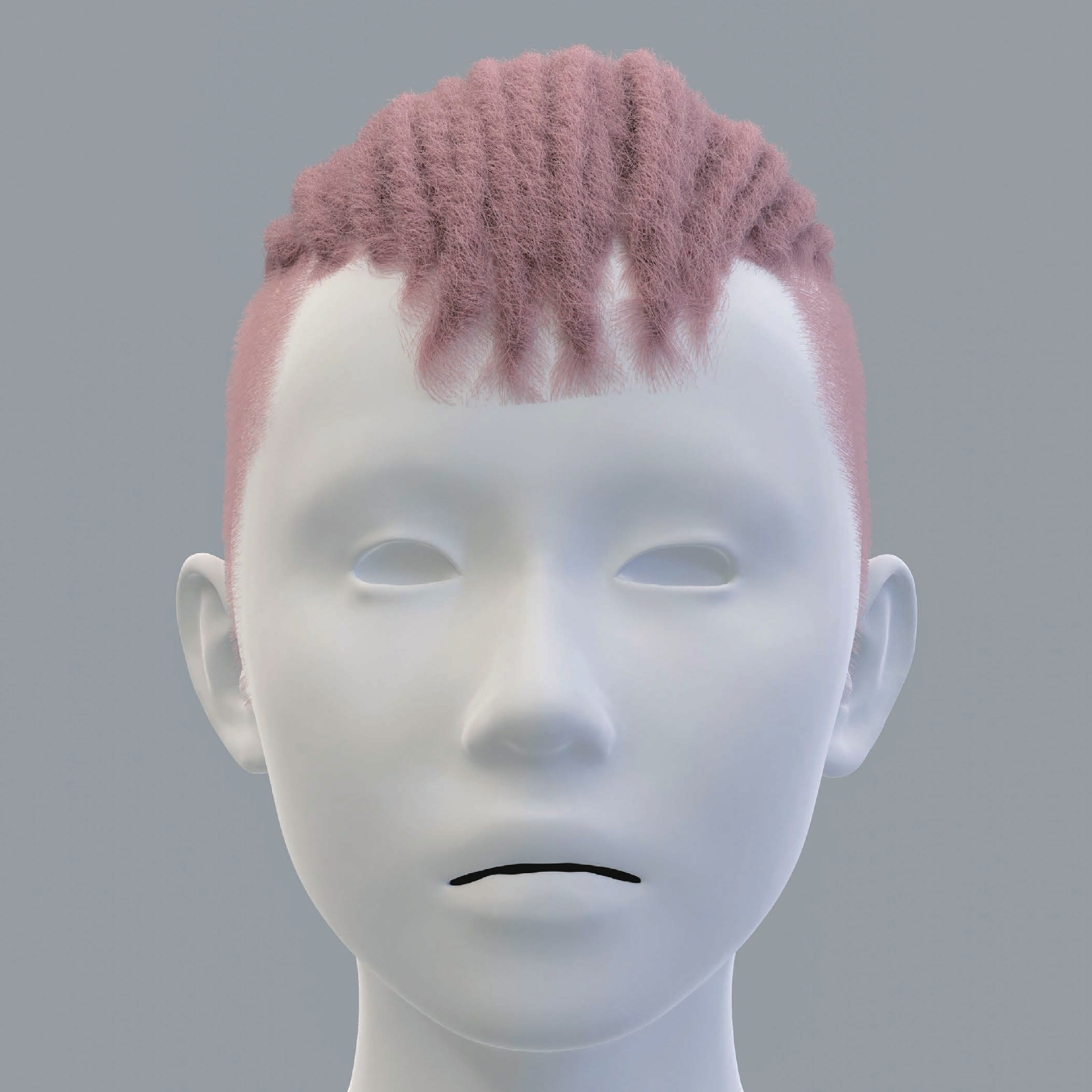}
\caption{(left) The source hairstyle and its transfer to another character with a large variance of pose, where artifacts arise. (right) Drastic hairline edits lead to unnatural distortions.}
\label{fig:limit_pose_change}
\end{figure}

\section{Conclusion, Limitation, and Future Work}
We present a method to automatically adapt a 3D hairstyle from the source character to a distinct target, and we design a constrained optimization framework to achieve this. With dedicated energies and constraints, the adapted 3D hairstyle can faithfully preserve the shapes and spatial interactions of the source models. Considering the high-resolution nature of 3D hairstyles, we build a two-level hierarchy for strands and employ a multi-scale solving strategy that substantially accelerates the nonlinear optimization. Moreover, to satisfy the user requirements for hairline edits, we apply an embedded membrane formulation to redistribute all the hair roots uniformly. The experiments demonstrate the effectiveness of our method.

Our method requires the same topology for the source and target characters, which is necessary for transferring the spatial positioning conditions of the hair-body relationship. However, algorithms like \cite{bogo2014faust,ovsjanikov2012functional} could be conducted first to build a dense correspondence when the character models have different topologies.

We also require the source and target characters to have similar poses. Although the optimization objective is pose-variant, our method is insensitive to a minor difference in pose. However, when significant pose differences arise (see Figure \ref{fig:limit_pose_change} left), the physics-based hair simulation might be feasible to produce visually plausible results \cite{daviet2011hybrid,kaufman2014adaptive}.

The quality of hair-root redistribution could deteriorate under a drastic hairline edit (see Figure \ref{fig:limit_pose_change} right). In such a case, the texture synthesis technique for hair, e.g., \cite{wang2009example}, is a better choice.


Due to its robust and effective retargeting capability, our method can be regarded as a data augmentation tool to increase the variances and capacities of existing hair datasets, e.g., \cite{hu2015single}. Training neural networks on such augmented data is beneficial to advance interactive VR techniques for digital humans. Moreover, making our method differentiable could provide a semantic loss to enable self-supervised learning.

\ifCLASSOPTIONcompsoc
  \section*{Acknowledgments}
\else
  \section*{Acknowledgment}
\fi

The authors thank the reviewers for their invaluable comments. This work is partially supported by NSF China (No. U23A20311, 62172363).

\ifCLASSOPTIONcaptionsoff
  \newpage
\fi



%
\bibliographystyle{IEEEtran}
\bibliography{./reference}

%

\begin{IEEEbiography}[{\includegraphics[width=1in,height=1.25in,clip,keepaspectratio]{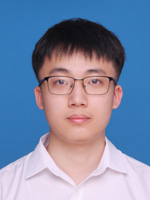}}]{Lu Yu}
received his bachelor’s degree in Applied Mathematics from Zhejiang University in
2022. Currently, he is working toward a PhD
degree at the State Key Lab of CAD\&CG, Zhejiang University. His research interests include
geometry processing and physics-based simulation.
\end{IEEEbiography}

\begin{IEEEbiography}[{\includegraphics[width=1in,height=1.25in,clip,keepaspectratio]{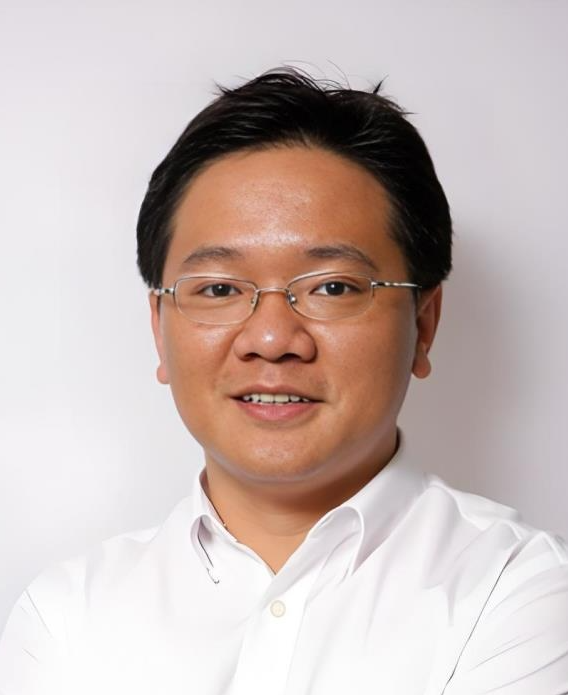}}]{Zhong Ren}
received his bachelor's degree in Information Engineering, his master's degree in Mechanical Engineering, and his PhD in Computer Science, all in 2007 from Zhejiang University. He is currently an associate professor at Zhejiang University and a member of the State Key Laboratory of CAD\&CG. Before returning to Zhejiang University in 2010, he spent three years at Microsoft Research Asia, where he served as an associate researcher in the Internet Graphics Group. His research interests include real-time rendering and GPU-based photorealistic rendering.
\end{IEEEbiography}

\begin{IEEEbiography}[{\includegraphics[width=1in,height=1.25in,clip,keepaspectratio]{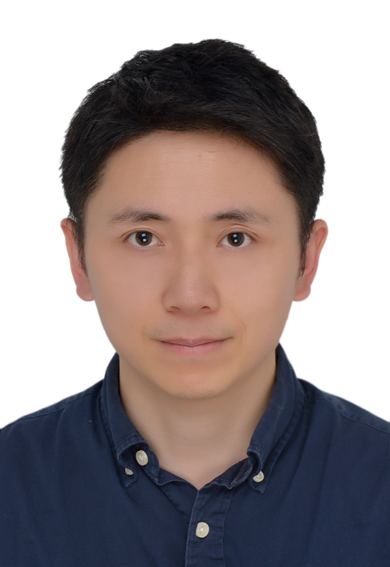}}]{Youyi Zheng}
is a ZJU100 Young Professor at the State Key Lab of CAD\&CG, College of Computer Science, Zhejiang University. He obtained his Ph.D. from the Department of Computer Science and Engineering at Hong Kong University of Science and Technology, and his M.Sc. and B.Sc. degrees in Mathematics, both from Zhejiang University. His research interests include geometric modeling, imaging, and human-computer interaction.
\end{IEEEbiography}

\begin{IEEEbiography}[{\includegraphics[width=1in,height=1.25in,clip,keepaspectratio]{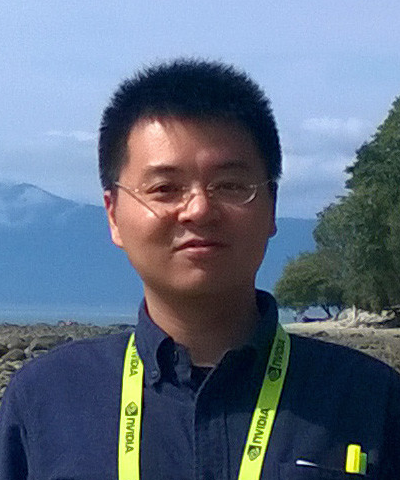}}]{Xiang Chen}
is an Associate Professor at the State Key Lab of CAD\&CG with Zhejiang University. He received his BSc and PhD both from the department of Computer Science at Zhejiang University. His research interests include geometry processing, physics-based simulation, data analysis, and computer-aided design.
\end{IEEEbiography}

\begin{IEEEbiography}[{\includegraphics[width=1in,height=1.25in,clip,keepaspectratio]{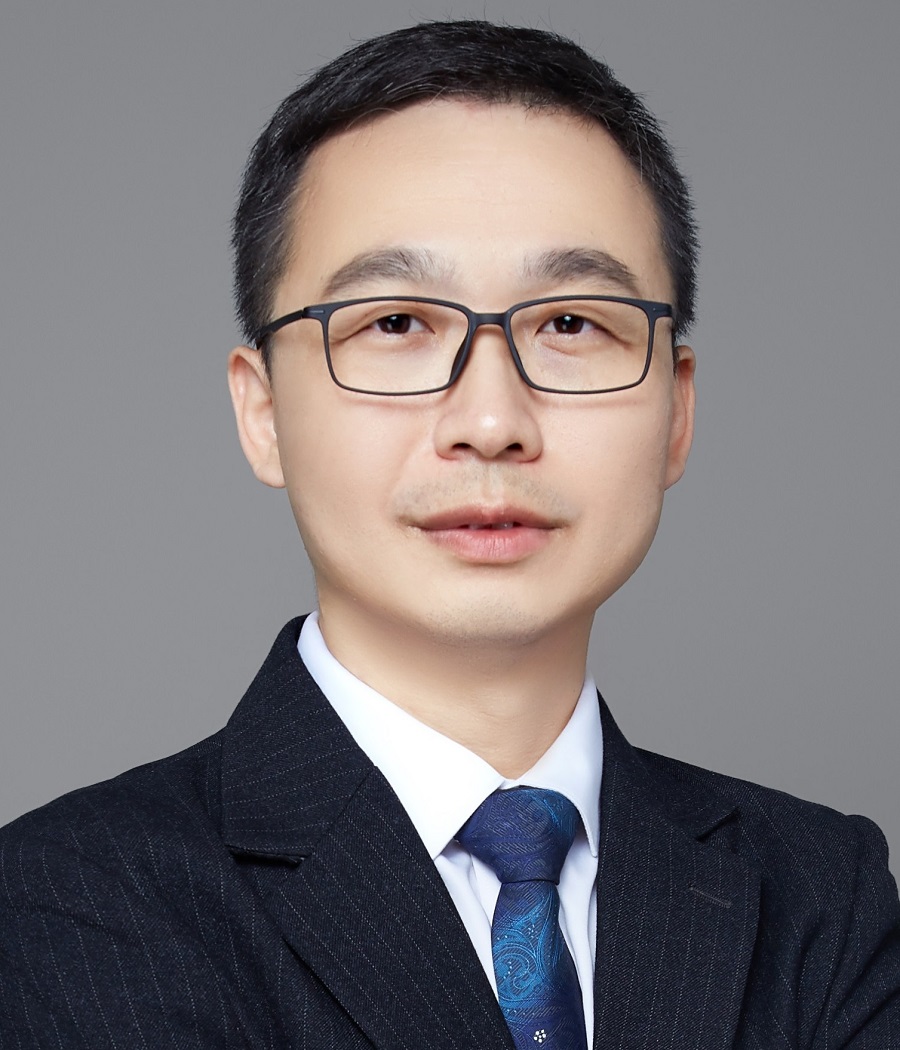}}]{Kun Zhou}
is a Cheung Kong Professor in the Computer Science Department of Zhejiang University and the Director of the State Key Lab of CAD\&CG. He received his BS degree and PhD degree in computer science, both from Zhejiang University. After graduation, he spent six years with Microsoft Research Asia and was a lead researcher of the graphics group before moving back to Zhejiang University. He was elected as an IEEE Fellow in 2015 and an ACM Fellow in 2020.
\end{IEEEbiography}

\vfill







\end{document}